**Bi-Demographic Changes and Current Account using SVAR Modeling: Evidence from Saudi Arabia**

Hassan B. Ghassan [1]    Hassan R. Al-Hajhoj [2]    Faruk Balli [3]

Version 2

[1] Corresponding author (Umm Al-Qura University),    hbghassan@yahoo.com
[2] King Faisal University, School of Business,    hajhouj@kfu.edu.sa
[3] Massey University, School of Economics and Finance. F.Balli@massey.ac.nz

**Abstract**

The paper aims to explore the impacts of bi-demographic structure on the current account and growth. Using a SVAR modeling, we track the dynamic impacts between these underlying variables. New insights have been developed about the dynamic interrelation between population growth, current account and economic growth. The long-run net impact on economic growth of the domestic working population growth and demand labor for emigrants is positive, due to the predominant contribution of skilled emigrant workers. Besides, the positive long-run contribution of emigrant workers to the current account growth largely compensates the negative contribution from the native population, because of the predominance of skilled compared to unskilled workforce. We find that a positive shock in demand labor for emigrant workers leads to an increasing effect on native active age ratio. Thus, the emigrants appear to be more complements than substitutes for native workers.

**JEL Class**. C51, F22, F41, J15, J23.

**Highlights**
- We explore the impacts of bi-demographic structure on current account and growth
- New insights are developed on the interrelations between bi-population, CA and growth
- The demand labor for emigrants has a positive long-run impact on economic growth
- A positive shock in emigration has a positive impact on native active age ratio
- The emigrant workers appear more complements than substitutes for native workers

## 1. Introduction

The macroeconomic literature, analyzing the effects of aging and demographic structures especially on the current account, have regained increasing interest among economists and politicians. This issue is growing importance, due to both the advanced and emerging economies dealing with an aging population problem in both the short and long term. According to the life cycle hypothesis (Modigliani and Ando 1957, Ando and Modigliani 1963, Modigliani 1966), individuals manage variability in their consumption and savings during their life-cycles. This hypothesis predicts that people in their working age consume a smaller fraction of their current income, compared to younger and older people. Dynan et al. (2009) analyses explicitly the effects of the demographic structure on consumption and saving behaviors. They find that an increase in the middle-age (30s-50s) household-group tends to save relatively more than other ages-group, when their income peaks. However, an increase in the young and old-aged households tends to dissave; in other words, the young households borrow against expected future earnings, while senior households draw down their accumulated savings. A decline in child population, caused by a progressive decreasing in the birth rates, will reduce aggregate demand for goods and investments. Since the current account balance (CAB) is a net picture of the interactions between the consumption and investment behaviors during the life-cycle processes, it is affected by the demographic change and the population structure.

In this paper, our contribution to the existing literature is to consider among traditional factors, the impact of bi-demographic structure (native citizen and emigrant resident) on the current account.[1] We will measure the net impact of the bi-demographic composition (citizen versus expatriate) on the current account balance and economic growth by dropping (keeping) the emigrants' population. Such modeling allows to empirically explore the evaluation of the group-age effects of expatriates and citizens on the current account balance of the Saudi economy.

---

[1] This new term of bi-demographic structure is justified by the intensity of the emigrant size to the native population. In 2016, the emigrant working age population was about 8.8 million which is very close to the native active population of 9.4 million. The total emigrant population was around 11.68 million, and the total of native population was about 20.06 million. In 1975, there was a large difference, the census of 1974 indicates that the expatriates of 25-64 age are about 0.35 million and the corresponding native citizens are around 1.9 million (see Figures C.0, Appendix C: Population pyramid).

According to a well-documented literature survey of Hassan, Salim and Bloch (2011), when the life cycle hypothesis (LCH) is extended to open economies, it implies that age composition affects the current account balance, with a positive effect of the working-age population and a corresponding negative effect of the young dependents population.[2] They indicate that there is a lack of a formal theoretical model, and that this area of research is at its first stage. More evidently, the impacts of the hybrid demographic structure on the current account in developed and developing countries have not been yet explored in related literature. This paper examines such issues and contributes to the underlying literature by splitting up the population into the natives and emigrants to check their respective effects on current account and economic growth.[3]

We find that a positive shock in domestic active age ratio contributes positively to the current account balance and increases the economic growth marginally. This means that the Saudi economy undergoes a fundamental transition in demographic and economic variables. By compiling demand labor for emigrant and native working-age population shocks, we detect that the net effect on the current account is positive indicating a tendency predominance of skilled emigrant workers. Further, the findings show that the positive shock in the migration flows leads to a negative impact on native active age ratio. Consequently, the emigration flow can reduce the labor market prospects of natives for some specific skills. According to our empirical outcome, the emigrants appear to be for natives i.e. Saudi workers more complements than substitutes. In Section 2 we review the literature discussing the relationship between demographic changes, current account balance and economic growth. Section 3 displays the demographic and economic stylized facts of Saudi Arabia. Section 4 presents and discusses the time series of empirical model and its estimation. The results will be discussed in Section 5 and we conclude in Section 6.

---

[2] Initially, the life cycle hypothesis of consumption and saving behaviors is a microeconomic approach related to the age profile of an economic agent. The analysis is extended to population structure using the overlapping generations framework, which supposes that the representative agent lives for four periods: childhood, young working age, old working age and retirement.

[3] The term emigrant is more appropriate in our case because the vast majority of expatriates has restricted visas and are connected to the persons or companies where they work through a sponsorship system (named Kafala i.e. cautioner system) by which they renew their residence (Iqama) annually i.e. temporarily residents and under a strict control of the cautioner system.

## 2. Literature review

The pioneer papers addressing demographic factors focused on the analysis of the effects of the saving rates. By analyzing saving rates in a large cross-section sample, Leff (1969) finds a negative relationship between the dependency rate and aggregate saving rate.[4] Assuming the life-cycle hypothesis, Fry and Mason (hereafter FM, 1982) state that the saving rate depends not only on the global dependency rate (level effect), but also on the interaction between the youth-dependency ratio and GDP growth (growth-tilt effect). They suggest, in the steady-state population equilibrium, that the real rate of return on financial assets and the dependency rate determine the timing of saving through the life cycle. Using a panel data for about 50 countries, they find empirical evidence for such model by isolating a negative relationship between youth dependency and saving rates, and after controlling for the interactive effects of dependency and income growth.[5]

Taylor and Williamson (1994) apply FM's model (1982) to examine over a century of savings behavior in Canada, Australia and Argentina. They find suggestive support of demographic origins for late 19th century capital flows. In a partial equilibrium approach, Taylor and Williamson (1994) assume a full pass-through of surplus savings into the current account, and then overestimate the impact of dependency rate on current account balance. Taylor (1995) also applies FM's model to Latin American savings and investment behaviors and determines its implications for the evolution of the region's current account balances during the first decades of the 20$^{th}$ century. The finding of Taylor (1995) indicates evidence that the dependency rate is a significant variable of capital flows. By identifying a significant statistical relationship between demographic structure and the current account-to-GDP ratio and based on lower dependency rates, Taylor (1995) expects that demographic changes in Latin America region, except Venezuela as an oil exporter economy with current account surplus, could reduce its current account deficit through encouraging savings processes during mid-life years. Higgins (1998) explores the relationship between age structures, saving, investment, and current account balance using panel of 100 countries. He documents that increases in both youth and old-age dependency have negative effects on savings and investment rates, and thus have a role in determining the current account balance. Higgins (1998) points out that a lower dependency rate implies a current account surplus. Using an overlapping generations model, Brooks (2000) analyzes the cohort growth and productivity growth effects on the CAB. He indicates that differences in cohort population growth around the world determine the international capital flows. He finds that globally falling (rising) in cohort growth would generate current account surplus (deficit) and capital outflow (inflow).

---

[4] "Dependency rate" refers to the ratio of children (below 25 years) and old aged (above 65 years) to the working age population (from 25 to 65 years). Other papers, like Fair and Dominguez (1991), use the active age structure ratio as the inverse of the age dependency rate.

[5] The demographic effects on savings have been analyzed by a different approach using data from the National Family Income and Expenditure Survey and leading to contrasting findings. For instance, Deaton and Paxon (1997) find that demographic change has a modest impact on savings because there is co-residence effect implying that older and younger adults share the same household.

Feroli (2006) examines whether the old-age dependency rate in the US trading partners has been a significant cause of the long-run capital flows to the United States. He simulates a multi-region overlapping generation model to match age structure differences between japan and North America. Feroli (2006) documents that aging is occurring rapidly in Japan compared to North America and indicated that the demographic changes play a significant role in determining the size of US current account deficits. Similar results are found in the paper of Kim and Lee (2007), which considers a panel data set of 10 East Asian countries over 1981-2003. They conclude that the speed increase in dependency rates in East Asian countries would have significant negative impacts on saving rates, global capital flows and current account imbalances. Also, Kim and Lee (2008) consider the effects of demographic change on the current account balance in the group of seven countries (G7). They find that total dependency rates deteriorate the current account balance and that an increasing rate of aging population might cause, in many of the G7, a decline in saving rates and consequently a deterioration in the current account balance. Karras (2009) examines the relationship between population growth and national saving, investment, and current account imbalance, using panel data covering the period from 1970 to 2003 and samples of 154 countries. He uses the overlapping generation model to estimate these relationships. Karras (2009) finds that saving and investment rates are negatively related to the population growth. However, he indicates that these effects on saving are stronger, and consequently, the current account is negatively affected by both the population growth and the size of the government sector.

The empirical results exhibit no consensus in the sign of demographic impacts on saving and investment. Li et al. (2007) consider the per capita impacts of both old-age dependency and longevity on savings, investment, and economic growth rates using panel of 200 countries from 1960 to 2004. They find that both saving, investment and economic growth are positively related to longevity, while their effects are negatively related to the dependency rate. The results also indicate that both population age structure and life expectancy are significant factors in determining the economic growth. Zheng (2007) explores the influences of demographic transition on saving rates in different regions. He divides the dependency ratio into youth and old-age dependency ratios. The results show that the increase in youth-age dependency ratio has a negative effect on savings rate, while the increase in old-age dependency ratio has a positive effect on the savings rate. Cooper (2008) argues that youth and old-age populations could have opposite effects on the current account balance. As the young people will be part of the labor force in the future, the current investment will be attracted by countries with large young-age populations than countries with an old-age populations and low birth rates.

Marchiori (2011) examines the impact of demographic changes on international capital flows and current accounts. He builds a multi-region overlapping generation model and uses data over the 21$^{st}$ century. The results are consistent with the life cycle hypothesis. Graff et al. (2012) examine the impact of demographic changes on the current account using a panel of 84 countries. They indicate that people may save more when they move toward retirement. Meanwhile, people retirees are more likely to dissave due to longevity uncertainty. They conclude that population aging does not have discernible impacts on the current account balance.[6] Bjorvatn and Farzanegan (2013) examine the interaction effect between resources wealth and demographic changes on the economic development using a panel of 120 countries. They find that an increase in natural resource rents reduce the income effect of demographic changes. Recently, Gudmundsson and Zoega (2014) explore the effects of population age structure on the current account imbalances for 57 countries over 2005-2009 by adjusting the current account surpluses and deficits considering the differences in the age structure of the population. They indicate significant effects of age structure on the current account balances. Fukumoto and Kinugasa (2017) explore the relationship between age structure and trade openness using panel data of 85 countries from 1991 to 2010. They indicate that an increased percentage of the working-age population in a total population has a positive impact on trade openness, while an increase in the share of a dependent population has negative impact on trade openness. They argue that this result may reflect the fact that a dependent population tends to require more spending than working-age population of non-tradable goods.

---

[6] By considering less parsimonious models, Graff et al. (2012) accept the negative effect of youth-age dependency on the current account but the effect of the old-age dependency is not empirically evident. The additional explanatory variables in the current account regression, such financial openness and natural resources rents, complicates the plausibility of the regression outputs since there will be a severe problem of multi-collinearity invalidating the statistical significance of the key demographic parameters.

When the economy needs more active population for some specific skills that couldn't be satisfied in the short-term by the native population, the international emigration could lead to adjust the disequilibrium in domestic labor market (Dustmann 2005). The intensity of the demand for emigrant workers by private and public sectors depend on the magnitude of the domestic labor market imbalances.

By synthesizing the literature review, there is no consensus about the impacts of age structure, in terms of age dependency ratio, on the current account and growth. The differences between the intensity and the direction of demographic factors between economies depend on their main features as well as on the preferences and behaviors of their agents during a certain temporal horizon. But, overall it appears that there is a negative relationship between demographic dependency rates and the current account (Leff 1969, Taylor and Williamson 1994, Taylor 1995, Higgins 1998, Feroli 2006, Kim and Lee 2007, Kim and Lee 2008, Karras 2009). Zheng (2007) and Copper (2008) distinguish between the negative effect of youth-age dependency and the positive effect of old-age dependency ratios on the current account balance, and the net effect depends on the intensity of these two opposing forces. On the other side, there is a sensitive relationship between economic growth and the current account balance (Fry and Mason 1982, Li and al. 2007). By distinguishing between native-born emigrant and foreign-born skill levels from OECD panel data Boubtane et al. (2013) find that there exists a positive impact of emigrants' human capital on economic growth. The analysis of Brooks (2000), about the determinant effect of cohort population growth on CAB and capital flows, can extended to income and capital outflows by postulating the importance of the bi-cohort active population growth effect on the CAB. However, there is no research paper addressing the demographic effects of native and emigrant age structure on specifically the current account or generally on macro variables. This paper contributes to fill this gap in the related literature by modeling explicitly the impacts and shocks of the native and emigrant dependency rates on the current account balance and economic growth.

## 3. Economic and demographic data and stylized facts of Saudi Arabia

Many countries have experienced substantial demographic change during recent decades, but the particularity of GCC countries, as in Saudi Arabia, is that the increasing number of expatriates has transformed the age pyramid of the bi-population (i.e. native and emigrant) since the 1970s years. From the official censuses data, the Figures C.0.1 and C.0.2 (Appendix C) of population age structure show that the difference in the proportion of working age between the expatriate and the Saudi was less than 30% until 1980, but in 1989 this gap reached its maximum around 39% (Figure 1a). The 2010 and 2016 censuses exhibit an increase in the working age population for both the natives and expatriates, and mostly for natives. From 1974 to 2016, via the mechanisms of the overlapping between generations, Saudi young group-age 00-24 tends to decrease continuously, thus leading to an increase in the Saudi working age (Appendix C). The proportion of expatriates 25-64 working age becomes more important compared to its corresponding Saudi working age (Figure 1a), but a negative trend starts from 1998.

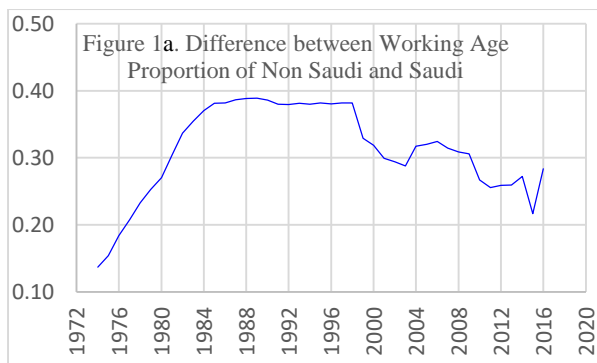
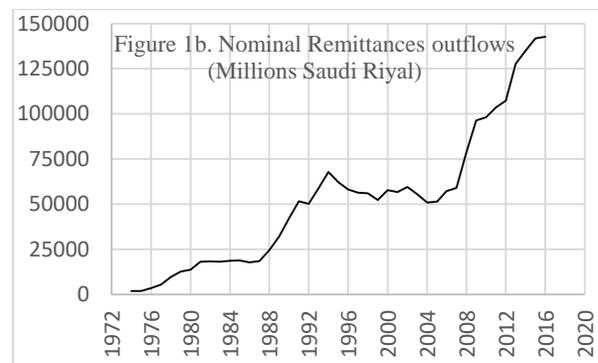

Figure 1a. Difference between Working Age Proportion of Non Saudi and Saudi

Figure 1b. Nominal Remittances outflows (Millions Saudi Riyal)

Domestic factors such as a qualified native youth population, in addition to the housing strategies and the restricted emigration to the qualified workers, could contribute to increase the surplus or reduce the deficit of CAB through the overlapping generation processes. The structural slowdown in the growth of global trade, causing more competitiveness for exporting crude oil and petrochemical products, makes the domestic factors more crucial. The correlation between economic variables and the changes in the demographic age structure is scarcely used in empirical papers.

The evolution of the correlation between CA-to-GDP and bi-population age structure is outlined in Figure 2. Basically, by considering the correlation between CA-to-GDP ratio and the hybrid age structure, the findings indicate that Saudis young 0-24 age is negatively correlated to CAY ($-0.348$ with Probability-value $0.011$). While both Saudis old 65+ age and Saudis working 25-64 ages are positively correlated to CAY, the results are $0.316$ with P-value $0.019$ and $0.334$ with P-value $0.014$, respectively (Tables 1 in Appendix B).[7] The sign correlation of Saudis old 65+ and CAY does not corroborate with the conventional life cycle hypothesis. This result can be explained by the specific social organization related to religious culture and Islamic belief of Saudi citizens who live in a common familial housing, and by the specific economic management dominated by the familial enterprises.

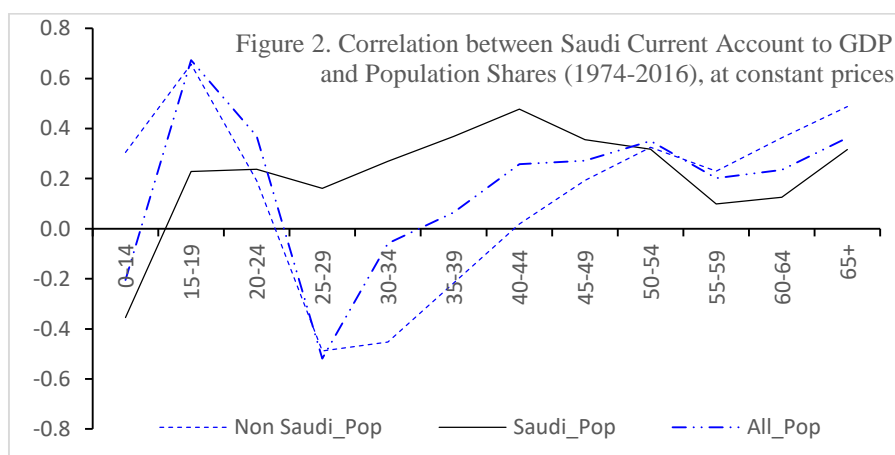

Figure 2. Correlation between Saudi Current Account to GDP and Population Shares (1974-2016), at constant prices

More interesting are the outcomes of the non-Saudi population. We find in contrast to Saudis working age that the correlation between emigrant working age and CAY is negative ($-0.394$ with P-value $0.0045$). This means that the main component of the emigrants in working age directly affects the CAB negatively through the remittances (Figure 1b), but the added value generated by the emigrants' contribution in economic activities would be largely profitable for the host economy (Peri 2016).[8] The emigrants obtain lower wages allowing the firms to save more money, in addition to their productive externalities by increasing the density of economic activities. This statement is confirmed by using the impulse responses analysis (See Sub-Section 4.5 and Figures C.1) and shows that the simple correlation remains a basic descriptive analysis and cannot capture the dynamic between economic and demographic variables.

---

[7] The proportion of age structure are calculated within each group of the expatriates and citizens.
[8] In Saudi Arabia as in the GCC countries, there are many emigrant workers in all sectors of the economy. According to CDSI and the ministry of Labor (2015) the emigrant labor force in the private sector reaches in average 87% during the last decade from 2005 to 2014; while in the government sectors, this average is around 8% (CDSI and the ministry of Civil Service 2015).

The question that arises is to understand if the remittance outflows can damage the competitiveness of the host country in the world markets. The increasing of remittance outflows would contribute to reduce the current account surplus or to rise the current account deficit of the host economy. The majority of emigrants are low-to-medium-skilled labor, leading to low wages.[9] But, by allowing comparative advantage in producing and exporting tradable items, the emigrants generate more net gains for the employees in comparison to their weak aggregated income; in the 2016 Census of population the emigrants represent 58.2% of the native population and the remittance outflows to GDP ratio is only about 5.8% (Figure 3b).[10] The majority of the emigrants lives without their family in the host country, and must remit a considerable share of the labor income to their family living in home country (Figure 1b). Consequently, at the macro level the Saudi host economy benefits more than the total income of emigrants. Considering the specific features of the economies as in GCC, we can theoretically state that there is no Dutch disease of the remittance outflows negatively affecting the current account balance.

As indicated by Castles and Miller (2009), Naufal and Genc (2012) and Amuedo-Dorantes (2014), the remittances are drives by many motives essentially the social motives. Nevertheless, according to Stark (1991) there is no general theory of remittances; the studies analyze this phenomenon based on empirical research which leads to statistical and economic evidence. Theoretically, it is expected that the employment of emigrant workers has a positive long-run net benefits for all the host economy or the home economy. Vargas-Silva and Huang (2006) indicate that the remittance is sensitive to changes in the macro conditions of the host country than in the home country. But, such effects remain an empirical challenge for both remittance inflows or outflows. A large literature focuses on remittances inflows than outflows, and there are little studies on host Gulf Cooperation Council (GCC) countries. A literature review on remittance outflows is well-documented by Hathroubi and Aloui (2016). The behavior of remittances and its impacts on the current account remain understudied in the economic literature. In theory and at the aggregate level, we consider that the remittances are affected by the variability across economic ups and downs in the host and home economies.

---

[9] In the Saudi economy the emigrant workers are largely from Asia and constitute more than half of the total workforce. Their restricted visas are directly connected to the persons or companies where they work through a cautioner (Kafala) system by which they renew their residence each year.

[10] The Fact Book 2016 of the World Bank Group indicates that the remittances mostly originate from high-income countries. By using the recorded outflows in 2014, it shows that the United States is by far the world's largest source of emigrant worker remittances, Saudi Arabia ranks as the second largest, followed by the Russia, Switzerland, Germany.

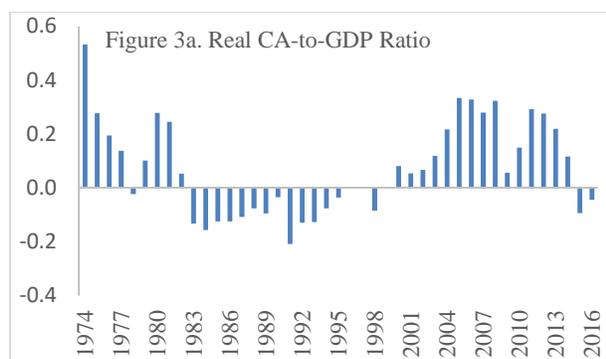

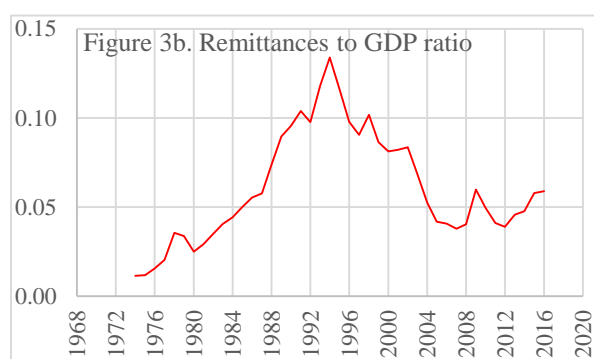

Such basic statistical outcomes appear to be in contradiction with the conventional LCH, but since the majority of expatiates must keep their family members in their native countries, they should remit money for their offspring's. Then, in the long-run, when we consider the rationale behavior of the emigrants, a bi-LCH would still be verified at least in GCC economies. This means that, by considering the bi-population behavior processes of natives and emigrants, there are two opposite cycles in the nexus between the current account and the population age structure (Figure 2).[11] Nevertheless, such explorative results need to be treated deeply by modeling the effects of dual-age structures on CAB. This should be done before confirming or invalidating the theoretical proposition that the bi-life cycle dynamics on CAB are the results of consumption and saving behaviors during the life cycles of successive generations.

---

[11] In this paper, we explore the empirical impacts of dual population age structure on the current account to GDP. But, this relationship must be discussed theoretically to determine the expected effects on the current account dynamic.

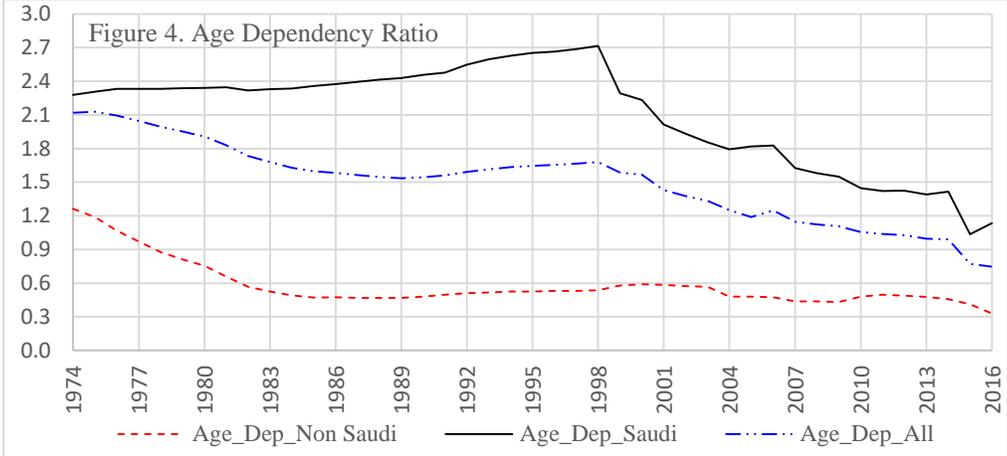

Such complicated dynamics of age structure effects on the CA-to-GP are definite challenges to economic and social policies, especially when the dependency rate of the Saudi population rises as before 1998 (Figure 4). As indicated by Simon, Belyakov and Feichtinger (2012), the age-specific emigration profile reduces the dependency rate. From Figure 4, it appears that the Saudi age dependency ratio is still greater than one, and the total dependency ratio becomes less than one due to the emigrant workers.[12] But, still, in Saudi Arabia that the total age dependency rate is higher about 75% in 2016, with an increasing trend of the native labor force. It is expected that economies similar to the Saudi economy, with a decreasing dependency rate and considerable natural resources, may potentially generate current account surpluses.

The organized social integration and assimilation of qualified emigrant workers in Saudi Arabia, as in many developed countries in occident would reduce the remittance processes.[13] The Saudi jobs are mostly higher-paid in the government sectors, but in contrast the non-Saudi jobs are largely lower-paid in the private sector. By eliminating this duality of the labor market in both public and private sectors, the wages will be more connected to the productivity of workers without considering their nationalities.[14]

---

[12] The effective expatriate dependency rate is not easy to calculate because it involves mostly the population age-structure of all the nationalities living in Saudi Arabia. Following the General Authority for Statistics (Demographic Survey 2016), we define the expatriate dependency rate by considering the emigrant residents in Saudi Arabia.

[13] Docquier, Rapoport and Salomone (2012) make an interesting discussion on such purpose by relating the remittances of expatriates to their educational levels. They indicate that the emigration policies in Gulf Countries increase the remittances,

As documented by Asharaf and Mouselhy (2013), the aging phenomenon occurs only to the native population but not to expatriates because the emigration policy is employment-oriented. Also, the emigrant workers are mainly male than female; most of them live without their families (See bottom of the Figures C.0.1 and C.0.2 in Appendix C). According to Asharaf and Mouselhy (2013), there is a slow but steady shift in age structure; the slowness is due to the influx of emigrants. In other words, there is a delay in the demographic transition process caused by many economic and religious factors.

## 4. Modeling

### 4.1. Unit root versus stationarity tests

Due to the underlying series related to population structure, current account to GDP and real economic growth over the period of estimation from 1974 to 2016, we consider that the Kwiatkowski, Phillips, Schmidt and Shin (KPSS) (1992) test is a valuable tool when looking for the presence or absence of random walk in such series (see methodological details in Appendix A.1).

In our empirical work, we consider values of the lag order truncation parameter $l$ from one to four, which is also used in the estimation of the long-run variance. The values of the test statistics are sensitive

---

while those conducted in European countries act to reduce it. To diminish the remittance outflows, we believe that emigrant human assets would be integrated and assimilated progressively, but the real productivity and cultural assimilation should be taken into account. Furthermore, to generate an anticipated net positive impact of the emigrant working age, it will be opportune to adjust the salaries of the emigrant workers to their realistic marginal productivity.

[14] According to CDSI (General Authority for Statistics, GaStat 2015) and Saudi Ministry of Labor (Labor market report July 2016, page 15), most emigrant workers are low-skilled (intermediate and primary education levels) and habitually employed with low wages in construction, retail and whole sale trade, personal services and manufacturing. More detailed information on the Saudi labor market can be found in the outlook and reports (2016) of Jadwa Investment (www.jadwa.com/).

to the choice of $l$; the value of KPSS decreases as $l$ increases.[15] This happens because $s^2(l)$ increases as $l$ increases, reflecting persistent positive autocorrelation in the series. In the presence of a temporally dependent series the errors $\varepsilon_t$ are not $iid$ under the null hypothesis, and as indicated by KPSS (1992, pages 173 and 175) to reduce the autocorrelation effect we use a reasonable large value of lag truncation parameter $l$. For $l$ around 3, there is reasonable power for $T$ in a sample range of 50 to 100 if $\lambda > 0.10$. But, when we take a larger value for $l$ to avoid also the size distortions, we face a drop in test power. Consequently, we must trade-off between correct size and power. Using such procedures, we can reject the hypothesis of level-stationarity for $g_{GDP}$ and the hypothesis of trend-stationarity for all the other series, namely all population-structure series and the current account to GDP. The Quadratic Spectral kernel (without truncation) is an asymptotically optimal kernel, and useful for estimating long-run variance and heteroskedasticity and autocorrelation consistent (HAC) covariance matrix (Andrews 1991, Kiefer and Vogelsang 2005). The Quadratic Spectral window leads to a higher power in small samples (Hobijn, Franses and Ooms 2004).

By employing the Quadratic Spectral kernel, with Newey-West automatic bandwidth selection (Newey and West 1994) and without pre-whitening the underlying series, we reject the null hypothesis of stationarity for all series with the probability of rejection at a nominal 5% level (Table 2 and Tables 3, Appendix B).[16] In addition by using the residuals $w_t$ from the equivalent model by taking an MA(1) process for the errors, we get exactly the same results that the first differences of the series are stationary (Tables 3, Appendix B), meaning that the original series are $I(1)$. Also, by applying a pre-whitening filter in the variance estimation using ARMA process, we can escape the upward size distortion through reducing the temporal dependence and the KPSS test performance would be enhanced.

---

[15] The results are available on request.
[16] The automatic bandwidth selection leads to a significant reduction in the size distortion of the test in the relevant case of a highly autoregressive process (Kiefer and Vogelsang 2005; Hobijn, Franses and Ooms 2004).

**4.2. Long-run relationship between bi-age dependency ratio and current account-to-GDP**

Following the nonstationary outcomes through the KPSS stationary tests, the bi-population variables and CA-to-GDP share common trend by moving together in the long-run. But, this result does not necessarily mean that these variables are related i.e. having a cointegration relationship. We should check whether the residuals from their regressions are nonstationary by using the test for null hypothesis of no cointegration (Engle-Granger 1987, Phillips-Ouliaris 1990) or the null of cointegration to control the probability of rejecting a valid economic model (Hansen 1992, Park 1992).[17] According to Ogaki 1993, the null of cointegration is harder to construct than the null of no cointegration. Also, according to Hansen (1990), the residual-based tests, such as the Phillips-Ouliaris test, have much better power than the Johansen and Stock-Watson tests.

Empirically, we are not interested by the short-run dynamics of the underlying variables because the age dependency rates dynamics are more manifested in the long-run. As a method for estimating a single cointegrating vector, the canonical cointegrating regression (CCR, Park 1992) model works with stationary transformations of the underlying variables to obtain asymptotically efficient least squares estimators by removing the long-run dependence between the cointegrating equation and stochastic regressors of innovations (see methodological details of CCR in Appendix A.2). It asymptotically eliminates the endogeneity bias in the cointegrating equation. These procedures lead to improved results in terms of size and power of the involved tests of cointegration, but depend on the quality estimates of the cointegrating relationships to get the most reliable test results. As in the KPSS test of stationarity, the empirical estimation of long-run covariance parameters uses the VAR pre-whitening method and the

---

[17] In practice, Park (1990) uses his CCR method to test for the null of stochastic cointegration by adding the trend and quadratic trend in the single cointegrating regression $y_t = \mu_0 + \mu_1 t + \mu_2 t^2 + \gamma' X_t + \varepsilon_t$. If $\mu_2 = 0$, then the test of cointegration can eliminate both the stochastic trends and linear deterministic trends.

Quadratic Spectral kernel with Newey-West automatic bandwidth parameter selection. Ogaki (1993) indicates that such techniques can considerably improve the small properties of CCR estimators.

By considering a regression explaining the real CA-to-GDP $y_t = cay_t$ as a function of the economic growth and age-dependency rates with $X_t' = (s\_adr_t, ns\_adr_t, g\_gdp_t)$ where $s\_adr_t$ and $ns\_adr_t$ stand for Saudi and non-Saudi age dependency rates, respectively; $g\_gdp_t$ corresponds to real economic growth. Empirically, the consistent estimation of the full long-run covariance matrix $\Omega$ is operated by centering the residuals $\hat{u}_t$, correcting for degrees of freedom and using the quadratic spectral kernel function which use Newey-West automatic bandwidth (3.7836) and automatic lag length (3.0) according to the data. The results for contemporaneous and full long-run variance-covariance matrices are as follows

$$\hat{\Sigma} = \begin{bmatrix} 0.0128 & -0.0022 & -0.0025 & -0.0008 \\ -0.0022 & 0.0236 & 0.0050 & 0.0006 \\ -0.0025 & 0.0050 & 0.0028 & 0.0004 \\ -0.0008 & 0.0006 & 0.0004 & 0.0016 \end{bmatrix}, \quad \hat{\Omega} = \begin{bmatrix} 0.0270 & -0.0165 & -0.0110 & -0.0021 \\ -0.0165 & 0.0873 & 0.0229 & 0.0029 \\ -0.0110 & 0.0229 & 0.0111 & 0.0017 \\ -0.0021 & 0.0029 & 0.0017 & 0.0021 \end{bmatrix}$$

The output of CCR estimation is

$$\hat{y}_t^* = 0.209 - 0.168\ s\_adr_t^* + 0.379\ ns\_adr_t^* + 1.392\ g\_gdp_t^*$$
$$\quad (0.087) \quad (0.004) \quad\quad\quad (0.005) \quad\quad\quad\quad (0.012)$$

The probability values between parentheses show the statistical significance of estimated parameters. Based on this canonical cointegrating regression, we test for cointegration using both the null of cointegration and the null of no cointegration. The former tests of Hansen (1992) with $LMc = 0.83$ at P-value 0.049 and Park (1992) with $\chi_2^2 = 8.04$ at P-value 0.017 reject the null hypothesis. Also, the latter ones of Engle-Granger (1987) with $\tau = -2.91$ at P-value 0.487, and Phillips-Ouliaris (1990) $\tau = -2.92$ at P-value 0.481 accept the null hypothesis.

Thus, the main outcome is that there is no cointegration between the current account to GDP, Saudi age-dependency rate, emigrant age-dependency rate and economic growth. The absence of cointegration can be explained by the differences in the intrinsic logic growth of each series, the limited sample size and probable non-linearity in the underlying series (Engle and Granger 1991). The non-evidence of cointegration, meaning that there is no loss of information in the long-run dynamic processes, makes it suitable to estimate a Structural VAR model in the first difference.

The domestic age structure is affected by the emigrants working age, which reduces the age-dependency ratio i.e. increase active age ratio of the host economy (Peterson 2017). Such processes influence the economic growth and consequently impact the current account balance through consumption and investment behaviors of both natives and emigrants. Based on the literature review synthesis, the economic analysis particularly the bi-life cycle hypothesis, and considering the empirically significant canonical regression between CA-to-GDP, age-dependency rates and output-growth, we postulate that (i) the increase in the native age dependency affects negatively the CAB i.e. $\frac{\partial cay}{\partial s\_adr} < 0$, (ii) the increase in the emigrant age dependency has a positive effect on the CAB i.e. $\frac{\partial cay}{\partial ns\_adr} > 0$, and (iii) the economic growth impacts positively the current account balance i.e. $\frac{\partial cay}{\partial g\_gdp} > 0$. The first relationship is theoretically and empirically supported in most papers cited above. In accordance with Fry and Mason (1982) and Li and al. (2007), the third relation states the positive effect of economic growth on the CAB. Using the assumption (iii) and in reliability to the human capital theory of Barro growth-regressions (1991), and if the main emigrant population is at working-age i.e. eligible for jobs, we can predict that the emigrant age-dependency ratio has a positive effect on economic growth, and consequently the relationship (ii) is justified because the ratio $\left(\frac{\partial cay}{\partial g\_gdp}\right)/\left(\frac{\partial ns\_adr}{\partial g\_gdp}\right)$ is expected to be positive.

### 4.3. SVAR model specification

The empirical modeling focuses on explaining the impacts of the hybrid demographic changes and economic growth variability on the Saudi the current account. Such complex and dynamic temporal interactions are explored through SVAR modeling. Based on explorative analysis of the correlation between bi-working age and the CA-to-GDP, we expect to verify the bi-life cycle hypothesis (Bi-LCH, see Section 3) through SVAR modeling. By assuming that the conditional expectations obey to a linear projection based on lags of the underlying stationary variables of the VAR model, they can be written as follows:

$$Y_t = cst + \sum_{i=1}^{p} A_i Y_{t-i} + \varepsilon_t \Leftrightarrow C(L)Y_t = \varepsilon_t \tag{1}$$

with $Y_t' = (dcay_t, ds\_adr_t, dns\_adr_t, dg\_gdp_t)$, $C(L)$ is the polynomial matrix of lag length, and $\varepsilon_t$ represents the reduced error term with $E(\varepsilon_t \varepsilon_t') = \Omega_\varepsilon$ and $E(\varepsilon_t \varepsilon_{t-i}') = 0$. In addition to our original dependent variable CA-to-GDP, we consider that the GDP growth and the expatriate dependency rate are determined endogenously. Also, since the young and old age population depend on the entire population age distribution and the life expectancy is influenced by the economic growth (Prskawetz et al. 2004), and in accordance with D'Albis and Collard (2013), it is plausible to assume that Saudi age-dependent share is determined endogenously. The VAR model can be well-arranged as a SVAR model by imposing parameter-restrictions on the matrices A and B of the following form (Amisano and Giannini 1997, Breitung et al. 2004):[18]

$$AY_t = cst_0 + \sum_{i=1}^{p} \boldsymbol{A}_i^* Y_{t-i} + Bu_t \tag{2}$$

where $\boldsymbol{A}_i^*$ is the matrix of structural coefficients and $u_t$ is the structural error or shock. This error is assumed to be a white noise process with zero mean and time-invariant variance-covariance matrix $\Sigma_u$. When the matrix $A$ is invertible, it allows to model contemporaneous relations among the variables of $Y_t$. By pre-multiplying with $A^{-1}$, $\boldsymbol{A}_i = A^{-1}\boldsymbol{A}_i^*$ for $i = 1,2,...,p$, we obtain the VAR equations and the relation between reduced and structural errors is $\varepsilon_t = A^{-1}Bu_t$ and its variance-covariance matrix is $\Omega_\varepsilon = A^{-1}BB'A^{-1'}$ by supposing that the shocks $u_t$ are orthogonal, we get: $\Sigma_u = I$.

---

[18] The matrices $A$ and $B$ are both unknown constraints and parameters based on economic analysis and economic hypotheses. They serve to shift from reduced errors to structural errors that have economic meaning.

This so-called AB-Model cannot be estimated without combining the restrictions on A and B that are consistent with a-priori theoretical expectations (Amisano and Giannini 1997). This way allows identifying economic and population shocks. The number of non-redundant elements of the variance-covariance matrix $\Omega_\varepsilon$ is $K(K+1)/2$, where $K$ is the number of variables in the VAR. Accordingly, we can identify just $K(K+1)/2$ parameters of the structural VAR. Since there is $2K^2$ elements in the matrices $A$ and $B$, the number of required restrictions to identify the full $AB$-model is $2K^2 - K(K+1)/2$ which is equal to $K^2 + K(K-1)/2$. If the matrix $A$ or $B$ is set to be the identity matrix, then $K(K-1)/2$ restrictions remain to be imposed. Regarding the underlying variables of the basic VAR model with $K=4$, twenty-two restrictions should be imposed on the matrices $A$ and $B$, which are nine zeroes in each matrix $A$ and $B$ and four unit-elements in the diagonal of $A$.

Since the Saudi working age population is increasing faster than the elderly population, in addition to the emigrant contribution to the working age mass, we expect that the net decrease of the dependency burden should be accompanied by more economic growth to contribute positively in the current account dynamics. Consequently, any structural shocks in national dependency burden, with a negative correlation to CA, will impact negatively the innovations of the CA processes. Bonham and Wiemer (2013) observed in China and other Asian economies that the falling dependency ratio facilitated a high saving rate which depends on economic growth and would positively impact the current account.

Expectedly, there exists a positive short-run response of CA-to-GDP to a favorable shock of falling burden of dependency. But, due the bi-population system leading to a dual labor market, another stimulus coming from the expatriate dependency rate will affect the CA dynamics. The expatriate dependency rate is governed institutionally and essentially by the demand of the labor market from the private and public enterprises of the host country, and also from the households (see Section 3 for more details on the duality of this specific labor market). The condition of the emigrant workforce depends on the economic activities and the growth of per capita income. Any shocks in demand labor, for any reasons, including its own productivity and wages levels, would affect the expatriate dependency rate.

We consider that the domestic dependency rate is governed by the social and economic policies which lead to a growing population size and a smooth demographic shift along the age structure. Any shocks happening in the domestic working-age population would impact the Saudi dependency rate, and accordingly affect the current account dynamics through consumption and investment behaviors.

Since that from the AB-model there is a relationship between the reduced innovations and the structural shocks,[19] we have to matching original VAR variables to the corresponding economic shocks i.e. as new set of variables. We assume that the reduced innovations associated to the current account are originated from international trade, income and transfers shocks labeled TIT. The reduced innovations of Saudi age-dependency ratio emanate from domestic working-age population shocks labeled SWP; whereas the emigrant age-dependency ratio is derived from the dynamic of demand labor for emigrant workers labeled DEL. The reduced residuals of real economic growth arise from the real domestic supply labeled DOS.

Based on the literature review synthesis and the theoretical and empirical evidence of the long-run relationships between bi-age dependency ratios, economic growth and current account-to-GDP, we define the following system (3), where the first equation supposes that the current account-to-GDP ratio innovation $\varepsilon_t^{CAY}$ is determined by the structural shocks of international trade, income and transfers growth through the parameter $b_{11}$. The second equation results from the bi-life cycle hypothesis i.e. bi-LCH, assuming that Saudi age-dependency ratio innovation $\varepsilon_t^{S\_ADR}$ is driven by both shocks in Saudi working-age population growth and in growth of demand labor for emigrants through the structural parameters $b_{22}$ and $b_{23}$. The third equation exhibits that the emigrant age-dependency ratio innovation is determined jointly by the innovation happening in current account-to-GDP ratio innovation and by the structural shocks occurring in the growth of demand labor for emigrant workers and in the real economic growth. The fourth equation shows that the innovation in real output growth is driven jointly by the innovations in current account-to-GDP growth and by structural shocks that happen in domestic supply growth.

---

[19] The structural shocks are conventionally hidden; therefore, for their identification, they need economic theory and a-priori economic reasoning in addition to a-priori restrictions.

Therefore, we can suggest the following relationship between the reduced innovations and structural shocks in the system (3).

$$\begin{bmatrix} 1 & 0 & 0 & 0 \\ 0 & 1 & 0 & 0 \\ -a_{31} & 0 & 1 & 0 \\ -a_{41} & 0 & 0 & 1 \end{bmatrix} \begin{bmatrix} \varepsilon_t^{CAY} \\ \varepsilon_t^{S\_ADR} \\ \varepsilon_t^{NS\_ADR} \\ \varepsilon_t^{G\_GDP} \end{bmatrix} = \begin{bmatrix} b_{11} & 0 & 0 & 0 \\ 0 & b_{22} & b_{23} & 0 \\ 0 & 0 & b_{33} & b_{34} \\ 0 & 0 & 0 & b_{44} \end{bmatrix} \begin{bmatrix} u_t^{TIT} \\ u_t^{SWP} \\ u_t^{DEL} \\ u_t^{DOS} \end{bmatrix} \quad (3)$$

where the coefficients $a_{ij}$ stand for the response of variable $i$ to an unexpected shock in variable $j$; the coefficients $b_{ij}$ are the response of variable $i$ to a structural shock in variable $j$. The estimation of this system depends on the parsimony of the parameters number and their statistical significance using a bootstrapping to get the reliable standard deviation of the parameters of AB-model:[20]

The AB-system of fourth equations shows that the identification of SVAR is achieved in the light of the plausible economic analysis, and by imposing a specific set of identifying restrictions named $a_A$ and $b_B$ on $A$ and $B$. These restrictions appear after defining the economic parameters as unknown constraints on both $A$ and $B$. The natural set of restrictions are only contemporaneous because in the SVAR modeling the lagged dynamics are unrestricted. The necessary condition requires that the number of identifying-restrictions on $A$ and $B$ is superior (overidentification) or equal (just-identification) to the number of required restrictions of the full AB-model i.e. $a_A + b_B \geq 2K^2 - K(K+1)/2$ (in our case, $13 + 9 \geq$ $32 - 10$). The sufficient condition consists to check the non-singularity of the related information matrix which is a function of the parameters to be estimated (Amisano and Giannini 1997).[21]

---

[20] The trade-off between the just-identification and the statistical significance of the parameters in AB-model leads to prefer an over-identification, which requires to test the null hypothesis of the restrictions validity using likelihood ratio (LR) statistic. This latter is asymptotically distributed as $\chi^2_{(q_u - q_r)}$ where $q_u$ is the number of restrictions under just-identification and $q_r$ is the number of restrictions under over-identification. In our empirical work, we find that $\chi^2_{(10-8)} = 0.0425$ with a P-value 0.979 meaning that there is no statistical significance between outcomes of restricted and unrestricted identification.

When the identification and resolution of the system (3) is done, the parameters of the AB-model could be estimated by minimizing the negative of the concentrated log-likelihood:

$$\ln L_c(A, B) = -\frac{KT}{2}\ln(2\pi) + \frac{T}{2}\ln|A|^2 - \frac{T}{2}\ln|B|^2 - \frac{T}{2}\text{tr}(A'B'^{-1}B^{-1}A\widehat{\Omega}_\varepsilon)$$

where $\widehat{\Omega}_\varepsilon = T^{-1}\sum_{t=1}^{T}\hat{\varepsilon}_t\hat{\varepsilon}'_t$ is the estimated variance-covariance matrix of the VAR residuals with $\hat{\varepsilon}_t = Y_t - \widehat{cst} - \widehat{A}Z_{t-1}$ where $\boldsymbol{A} = (\boldsymbol{A_1}, \boldsymbol{A_2}, \ldots, \boldsymbol{A_i}, \ldots, \boldsymbol{A_p})$ is the parameters matrix with an order $K \times Kp$ and $\boldsymbol{Z}'_{t-1} = (Y'_{t-1}, Y'_{t-2}, \ldots, Y'_{t-p})$.

### 4.4. SVAR outputs

Using the previous results of the canonical cointegrating regression and the stationarity tests, a Structural VAR model in first difference is more suitable to estimate. Following Juselius (2006), some outliers should be removed prior to the estimation. According to Nielson (2004), the removal of outliers is basically equivalent to including impulse dummies in the regression. To reduce the effects of outlier's observations on series features, we apply the program TRAMO using the linearized series in the VAR estimation (Gomez and Maravall 1996). Firstly, the reduced VAR is estimated, with one lag resulting from the selection criteria tests, using stationary series obtained by the first difference. The stability condition holds for $VAR(1)$ because all autoregressive characteristic polynomial roots lie within the unit circle. Also, the multivariate serial correlation of residuals, through Portmanteau tests, is implemented to verify whether the residuals support the white errors assumption. The results indicate significantly that there is no autocorrelation between the reduced residuals. The calculated residuals' correlation and variance-covariance matrices are shown in Table 4.1 and Table 4.2. (Appendix B).

---

[21] For more details on SVAR identification and related algebra see Lucchetti 2006.

It is also of interest to test the heteroskedasticity between variances of VAR residuals. The test of Brown-Forsythe (1974) provides a BF-statistic about 33.42 with a p-value of $6.54E-17$, meaning a strong evidence of no homoskedasticity. Since the values of residual covariances elements are seemingly low, we have to test whether $\Omega_\varepsilon$ is a diagonal matrix i.e. there is no correlation across VAR equations. The likelihood ratio (LR) test, or alternatively the Lagrange multiplier (LM), are running for the joint significance of off-diagonal of the residual covariances matrix (Greene 2012 page 604, Enders 2004).[22] The null hypothesis is that the off-diagonal residual covariance elements are equal to zero, which corresponds to $\text{diag}(\widehat{\Omega}_\varepsilon)$; empirically, we compute this covariance matrix with $(1/T)\hat{\varepsilon}'\hat{\varepsilon}$ from the OLS residuals of each individual VAR equations with stationary variables. These OLS estimates are equivalent to the maximum likelihood (ML) when we suppose that the residual covariance is diagonal, and the errors are normally distributed. The alternative hypothesis is defined by the unrestricted $\widetilde{\Omega}_\varepsilon$ estimated by ML method. This test is decisive because if the nullity of residual covariances is true, there is no need to implement SVAR modeling.

The null hypothesis $H_0: \sigma_{12} = \sigma_{13} = \sigma_{14} = \sigma_{23} = \sigma_{24} = \sigma_{34} = 0$ is tested against the alternative hypothesis $H_1: \sigma_{12} \neq \sigma_{13} \neq \sigma_{14} \neq \sigma_{23} \neq \sigma_{24} \neq \sigma_{34} \neq 0$ using $LR = 2(LL_U - LL_R) = -2(LL_R - LL_U)$ where $LL_U$ and $LL_R$ are the maximum values of the log-likelihood function with unrestricted model and restricted model, respectively. We can obtain the exactly equivalent value by basing the statistic of LR on the difference between the restricted covariances matrix and unrestricted covariances matrix as follows

$$LR = T\left(\ln|\text{diag}(\widehat{\Omega}_\varepsilon)| - \ln|\widetilde{\Omega}_\varepsilon|\right) = T\left(\sum_{k=1}^{K} \ln \hat{\sigma}_k^2 - \ln|\widetilde{\Omega}_\varepsilon|\right)$$

---

[22] The Lagrange multiplier statistic of Breusch-Pagan (1980) can be determined by $LM = T \sum_{k=2}^{K} \sum_{l=1}^{k-1} r_{kl}^2$ where $r_{kl}$ is the residual correlation coefficient between equations $k$ and $l$ defined by $\hat{\sigma}_{kl}/(\hat{\sigma}_{kk}\hat{\sigma}_{ll})^{1/2}$. The limiting distribution of this statistic is $\chi_q^2$ as for the LR statistic. The LM statistic is easier to calculate because it does not require the ML estimates of $\Omega_\varepsilon$. From the Table 4.1, we obtain that $LM = 20.33$ which is greater than the critical value of 12.59, leading also to reject the null hypothesis of diagonal covariance matrix.

The statistic LR is distributed following $\chi^2_q$ where the degree of freedom $q$ is equal to $K(K-1)/2$ restrictions on the covariance matrix, $K$ is the number of variables in VAR model. The log-likelihood value of the restricted model under the null hypothesis comes from the single estimate of the equations in the VAR model and is defined by:

$$LL_R = LL_{dcay} + LL_{ds\_adr} + LL_{dns\_adr} + LL_{dg\_gdp}$$

We get the log-likelihood value of the unrestricted model i.e. under $H_1$ from the VAR estimates. Using the optimal lag length 1, we obtain from log-likelihood values $LR = 2(389.5953 - 379.0804) \simeq 21.03$ and from the covariance matrices, the result is $LR = 42(29.9036 - 29.4029) \simeq 21.03$ ; the critical value at 95% level of significance and with a degree of freedom $q = 6$ is 12.59. The calculated values are greater than the critical value, so we reject the null hypothesis and then the covariance are not jointly zero i.e. the shocks in different equations are contemporaneously correlated. This statistical feature justifies implementing the SVAR model which take into account the contemporaneous effects between underlying variables.

According to Kruskal's theorem, the projection of $Y$ on $Y_{-1}$ (or on $Z_{-1}$ for more lags) coincides with the same projection under the inner product $\Omega_\varepsilon^{-1}$ (for more details on Kruskal's theorem see Zaman 1996, Theorem 1.7, page 18). By supposing that $Y_{-1}$ (or on $Z_{-1}$ for more lags) is invariant under the transformation $\Omega_\varepsilon^{-1}$, then $\hat{Y} = \tilde{Y}$ where $\hat{Y}$ is the OLS estimates and $\tilde{Y}$ is the GLS ones. In other words, the OLS and GLS will be the same if and only if the $K$ columns of $Y$ are a linear combination of exactly $K$ characteristic vectors of $\Omega_\varepsilon$ (Greene 2012, page 334). Hence, there is no additional efficiency by applying on VAR model the ML estimation instead of OLS by individual equation since each equation in the VAR system has the same exogenous or predetermined lagged variables and there are no restrictions on the parameters and/or variance-covariances. The use of ML estimates, by restricting the variance-covariances matrix to be a diagonal matrix, leads to marginally improve the efficiency gain in running the LR test, because the difference between ML-LR and OLS-LR statistics is about $-8.43751E-07$. Generally, by supposing that the VAR model is under restrictions, Kim (2014) indicates empirically that the efficiency gain of GLS estimates increases as the contemporaneous correlation among residuals increases.

### 4.5. SVAR estimation

By using the structural factorization, we derive the estimated shocks by maximum likelihood for the SVAR model from the estimated VAR residuals. The estimated structure of shocks is showed by the following system, where the values between parentheses correspond to their respective probability values which indicate the statistical significance of the estimates

$$\begin{cases} \varepsilon_t^{CAY} = 0.0959\, u_t^{TIT} \\ \qquad\quad (4.947E-20) \\ \varepsilon_t^{S\_ADR} = 0.0143\, u_t^{SWP} - 0.0053\, u_t^{DEL} \\ \qquad\quad (4.993E-20) \quad (0.0210) \\ \varepsilon_t^{NS\_ADR} = -0.0365\, \varepsilon_t^{CAY} + 0.0078\, u_t^{DEL} + 0.0023\, u_t^{DOS} \\ \qquad\quad (0.0033) \qquad (4.992E-20) \quad (0.0485) \\ \varepsilon_t^{G\_GDP} = -0.1424\varepsilon_t^{CAY} + 0.0385\, u_t^{DOS} \\ \qquad\quad (0.0214) \qquad (4.948E-20) \end{cases} \qquad (4)$$

The simultaneous relationship between the underlying variables, through the matrix $\Phi_0$, indicates firstly the negative correlation between the shocks in the emigrant population and native population as mentioned in Tables 1. This negative structural slope approximately $-0.0053$ explains in term of shocks that the emigrant and native workers are in the short-run more complements than substitutes.[23] Also, the instantaneous impact of economic growth on the demand for emigrants is positive but small. The immediate impact of a positive shock in international trade, income and transfer growth can lead to decrease the demand labor for emigrants approximately $+0.0035$ (see matrix $\Phi_0$). This means that an improvement in international markets encourage the economy to hire more emigrant workers.

---

[23] By definition, an increase in native (emigrant) active age corresponds to a decrease in native (emigrant) age-dependency-rate, and vise-versa (see Footnote 4).

The effects of the structural shocks on the variables of interests are dynamics and can be explored by using the impulse response analysis, which leads to measure the effects of a one-time shock. For this purpose, we can determine the partial multipliers as follows $\Phi_0 = \tilde{A}^{-1}\tilde{B}$, $\Phi_1 = \hat{A}_1\Phi_0$, $\Phi_2 = \hat{A}_1^2\Phi_0$, ... where $A, B$ are the estimates of the AB-model, and $\hat{A}_1$ is the estimated matrix of parameters from $VAR(1)$. Also, the total impact matrix i.e. matrix of long-run effects can be calculated from $\Psi_\infty = (I - \hat{A}_1)^{-1}\Phi_0$ (for more details see Breitung et al. 2004, pages 166-167 and Lütkepohl 2007). The results are as follows

$$\Phi_0 = \begin{bmatrix} 0.0959 & 0.0000 & 0.0000 & 0.0000 \\ 0.0000 & 0.0143 & -0.0053 & 0.0000 \\ 0.0035 & 0.0000 & 0.0079 & 0.0023 \\ 0.0137 & 0.0000 & 0.0000 & 0.0385 \end{bmatrix}, \Phi_1 = \begin{bmatrix} 0.0130 & -0.0102 & 0.0132 & 0.0112 \\ 0.0007 & 0.0140 & -0.0096 & -0.0030 \\ 0.0027 & 0.0004 & 0.0074 & 0.0026 \\ -0.0160 & -0.0025 & 0.0002 & -0.0164 \end{bmatrix}$$

$$\Phi_2 = \begin{bmatrix} 0.0000 & -0.0106 & 0.0167 & 0.0024 \\ 0.0003 & 0.0132 & -0.0132 & -0.0033 \\ 0.0024 & 0.0009 & 0.0068 & 0.0022 \\ 0.0050 & -0.0004 & -0.0005 & 0.0061 \end{bmatrix}, \ldots, \Psi_\infty = \begin{bmatrix} 0.1614 & 0.0069 & 0.1662 & 0.0595 \\ -0.0573 & 0.0308 & -0.2025 & -0.0630 \\ 0.0117 & 0.0246 & 0.0077 & 0.0037 \\ 0.0041 & -0.0059 & 0.0122 & 0.0302 \end{bmatrix}$$

The dynamic long-run relationships between the underlying variables, through the total impact matrix of the shocks i.e. $\Psi_\infty$, show the extent of the cumulative negative effect with $-0.2025$ between structural shocks in the growth of demand labor for emigrants and the responses of native population growth. This negative structural relation explains that the emigrant and native workers are in the long-run more complements than substitutes. As well, even in the long-run, the cumulative impact of demand for emigrants growth on the economic growth is positive approximately $+0.0122$, but the domestic working-age population growth shock is associated with a negative economic growth of about $-0.0059$. In other words, a negative shock of Saudi active age ratio leads to a decrease in the output growth, and vis-versa. In the long-run, the net impact on growth of the two sources of working-age population growth is positive due to the predominant contribution of skilled emigrant workers in comparison to the unskilled of the mass of emigrant workers. These findings make new insights in studying the relationship between bi-population growth and economic growth within the neoclassical theory of population. On the other hand, the cumulative long-run impact of a positive shock in international trade, income and transfer growth could increase the growth of demand labor for emigrants about +0.0117, meaning that a long-run improvement in international markets motivates the Saudi economy to employ more emigrant workers. In the next section, we will discuss the processes of the responses of the underlying variables to diverse structural shocks.

## 5. Results and Discussion[24]

Theoretically, the current account reflects economic, social and financial shocks from both native and emigrant sides. Generally, the existing empirical literature focuses only on the link between emigration and host economic activity without considering the impacts of the economic interactions between natives and emigrants on the current account and growth. In this paper, we contribute to highlight the effects of emigrant working force dynamics on the CA-to-GDP, native active age ratio, and economic growth processes. By using a specific structural vector autoregressive framework, we can evaluate the amplitude of the impacts related to shocks of international trade, income and transfers (TIT), Saudi working-age population (SWP), demand labor for emigrant workers (DEL), and domestic supply of output (DOS). These factors play determinant roles in driving the Saudi Arabia's current account balance growth. The SWP and DEL shocks are fundamentally heterogeneous. The first shocks are connected to labor supply from the domestic working-age population, but the second shocks are related to labor demand for contracted emigrants. These two sources of labor could be complements or substitutes (Grossman 1982) as a labor factor, but it remains that the fluctuation in macro supply of labor, thereby in employment, depends jointly on the real demand addressed to the economy and the technological skills in the productive sectors. The theoretical analysis of this question is very controversial, as well as the empirical analysis, constructed inside a specific theoretical framework, which could lead to different outcomes. The nature of the relationship between population growth, the current account and growth depends on specific conditions in the countries or regions, mostly the age structure of the population leading to different types of age-dependency rate (Peterson 2017).

---

[24] As indicated in Stock and Watson (1996, 2001), the impulse responses functions in SVAR modeling could be sensitive to changes in lag length, sample period and identification restrictions. We check the robustness of the IRF results by modifying the lag length of the reduced VAR, the outcomes show that the shapes, signs and magnitudes overall remain robust to the changes from 1 lag to 2 lags.

Positive shock of international trade and income movement leads to increased growth of the current account-to-output ratio in the short-run; this impact changes throughout the temporal horizon (see Figures C.1, Appendix C). There is a hump-shaped reaction of CA-to-GDP growth indicating an increasing process after three years, this means that the TIT shock is highly persistent during the five first years. This outcome can be due to a sticky behavior hypothesis in the international trade and income flows. In contrast, any negative shock in the international markets of goods, services, factor incomes and transfers will mostly speed up in the short-run the CA-to-GDP process. But, the international liquidity of the Saudi economy can help to reduce any current account deficit (Ghassan et al. 2013). Also, smoothing the consumption along the temporal horizon and in between generations would support in reducing any burden from discrepancies in the financial and real international markets.

Positive shock in the domestic working-age population growth decreases the growth of the CA-to-GDP ratio. This outcome corroborates with the findings of many papers (see last paragraph of Section 2). During the horizon of first eight years, the domestic working-age demographic shock exhibits a negative effect on the current account. On the other hand, a negative shock in domestic working-age population, meaning an increase in Saudi active age, would positively contribute in the medium term to improve the current account balance. This positive relationship is supported in other papers as the paper of Fukumoto and Kinugasa (2017) which considers the dependency age of the entire population, but without dividing it into natives and emigrants. It appears that during the first fifteen years that the effects of native and emigrant working-age shocks on CA-to-GDP and economic growth go in opposite directions. The Figures

C.1 show that the positive reaction of CA-to-GDP to demand labor for emigrant worker shocks dominates and cancels out its negative reaction to domestic working-age population shocks. As predicted theoretically in Section 4.2, in both short and long-run horizons, there is a positive contribution of the emigrant working-age population growth on the CA-to-GDP growth.

Also a positive real domestic supply shock has a permanent positive impact on the responses of CA-to-GDP ratio. This indicates that any positive shock in macro-investment and then in the output would positively affect the current account growth. But, as the Saudi economy is dominated by oil sector activities, in the short-run the shocks in the output growth generates small effects on the current account balance, but at the long-run horizon these effects decay very slowly (see Figures C.1).

Furthermore, during a generation i.e. a horizon of first twenty-five years, a positive shock in domestic working-age population growth generates a positive hump-shaped reaction of the emigrant age dependency ratio growth (Figures C.3). This means that there is a complex substitutability process between emigrant and native working-age population when the shocks emanate from natives working-age. This outcome in term of shocks indicates that there is no a real aging phenomenon for the native population, at least for the next few decades. According to Asharaf and Mouselhy (2013) this phenomenon does not occur for expatriates, we perceive that is because the emigration policy is employment-oriented. Overall, the hump-shape of the impulse responses function (IRF) of the emigrant age dependency rate reflects the magnitude in the dynamic interaction between emigrant population and domestic population. Whereas, the self-responses of emigrant age dependency ratio appear to be monotonic to demand labor for emigrant workers' shocks. The explanation of this effect can be due to the non-imperfect information between the emigrant working-age forces and the agencies that manage the demand labor for emigrant workers.

By considering the IRF of Saudi age dependency ratio, we find that during a generation there is a negative hump-shaped reaction to emigration policy shocks i.e. an increase in demand labor for emigrant workers (Figures C.2).[25] The reason can be due to a dynamic complementarity process between native and emigrant working-age population when the shocks emanate from emigrants working-age. This means that a unitary increase in emigrant active-age drives to a decrease in Saudi age dependency ratio which is equivalent to an increasing effect on the Saudi active-age which is expanded slowly throughout the temporal horizon. Then, the emigrants in working age flow can enlarge the labor market prospects of natives for some specific skills. The reason for such outcome is that the domestic families and family businesses still need emigrant workers for any type of jobs either in economic or social activities such as private drivers or house-keeping jobs or other skilled and unskilled jobs.

But, it would still be that such logical deductions are true in mean, because, firstly, the low-skilled labor of emigrants is dominant towards the highly-skilled, largely in private retail activities; secondly, Saudi workers are employed mostly in the public sector and banking and finance activities; thirdly, the levels of emigration are highly regulated by the Saudi government. According to our empirical outcome, the emigrants appear to be, for natives i.e. Saudi workers, more complements than substitutes. This finding is supported in the empirical work of Boubtane et al. (2013). According to the basic Saudi economic facts, the complementarity works mostly in the private sector where the emigrants are predominantly employed, but they appear more substitutes than complements in the public sector dominated by Saudi nationals (MGI Report, page 6, 2015). However, the emigrant inflows do not strongly affect the established native workers, but could affect potential native workers mainly during bad economic cycles.

In addition, during the first fifteen years of the horizon, the self-shock of demand labor for emigrants positively affects the growth of the emigrant working age dependency, but such positive impacts decrease continuously during the temporal horizon (Figures C.3). This evolution can be explained by the technological factor that progressively reduces the need of emigrant workers mostly the unskilled ones. By combining the shocks of demand labor for emigrant workers and of domestic working age population, the net effect from des impulse responses of emigrant and Saudi age dependency shows that the complementarity process between emigrant and native population dominates. This means that the human capital of emigrants can eliminate the negative impact of the increase of Saudi age dependency rate.

---

[25] In comparison to previous literature conducted at level country, Dustmann et al. (2005) finds that, overall and at macro level of the British labor market, the immigration does not have adverse effect on native employment.

A positive shock of Saudi active age ratio growth leads to a volatile marginal economic growth (Figures C.4). The reactions of economic growth unsteady move in opposite directions to shocks of native and emigrant working-age populations, but in the long-run both effects become slightly positive. In the short-run, a positive shock of demand labor for emigrant workforce generates volatile impacts on economic growth; but in the long-run horizon, the effect is positive and small (see Figures C.4). This result could mean that in the long-run, through the increased qualifications and mechanisms of learning by doing, among emigrants the skilled workers with elevated productivity, dominates the unskilled workers with lower productivity. Such empirical outcome establishes in different manner some previous empirical assertions as in Dolado et al. (1994) and Ortega and Peri (2009) and Peri (2016). Using the Solow growth model, augmented by emigrant human capital, Dolado et al. (1994) emphasize that the negative economic growth effects of emigration become less important the higher the emigrants' human capital. By using logit discrete-choice modeling, Ortega and Peri (2009) assert in 15 OECD countries that there is no evidence of crowding-out of natives by the emigrants as flow, and that emigration increases the GDP of the host country. In fact, the emigrants do not disadvantage the native per capita income; furthermore, and due to their contractual status, the emigrant workers cannot fully react to real wage decreasing compared to the native workforce.

Concerning the variance decomposition analysis using structural VAR factors, it measures the contributions of each source of shock to the forecasted structural error variance of each endogenous variable in the SVAR model over a temporal horizon. This analysis indicates the importance of specific structural shocks in explaining the variance of an endogenous variable. The variance decomposition displays that the structural disturbance in demand labor for emigrants explains around 69% of the fluctuations in the emigrant age dependency rate (see Table 5, Appendix B). Also, we detect that the structural error variance in domestic supply of output around 79% is mostly due to uncertainty in the growth equation. After 10 years, around 2.7% of variance in emigrant age dependency is due to shocks in current account-to-GDP ratio. As well, almost 13% of the error variance in economic growth forecasts is attributed to the current account-to-GDP ratio, whereas less than 2.5% and less than 5.5% are imputed to emigrant and native dependency age equations, respectively, to in explaining structural variances of growth. Interestingly, the shock in demand labor for emigrants explains approximately 33% of the changes in current account-to-GDP ratio, and around 69% of the variation in Saudi age dependency rate. But, the shock in the Saudi working-age population explains merely around 15% of the changes in the emigrant age dependency rate. In contrast, the shock in Saudi working-age population explains only around 3% of the fluctuations in CA-to-GDP ratio, and only around 23% of the variation in Saudi age dependency rate.

## 6. Conclusion

By specifying the emigrant active age ratio, we contribute to highlight the effects of the emigrant working age dynamics on the CA-to-GDP, native active age ratio, and economic growth processes. By using a structural vector autoregressive framework, we can evaluate the amplitude of the impacts of international and domestic shocks on the native and emigrant active age ratios, current account and economic growth.

The positive (negative) shock in domestic working-age population growth decreases (increases) the growth of the CA-to-GDP ratio. This finding is also documented in many papers, but without splitting up the population into natives and emigrants. In other words, a positive shock in Saudi active age growth contributes positively to the current account balance. Also, a positive shock in growth of demand labor for emigrants leads to quasi-permanent positive impulse responses of the CA-to-GDP growth. This means that the emigrant workforce contributes to increase the CA balance in both the short and long-run horizons. By compiling the demand labor for emigrant and native working-age population shocks, we find out that the long-run net effect is positive indicating the efficient contribution of emigrant population to the growth of CA-to-GDP ratio.

By considering the shocks of demand labor for emigrants, we detect that there is a negative hump-shaped reaction of the Saudi age dependency ratio growth. This outcome indicates that a unitary increase in emigrant active age leads to an increasing effect on the Saudi active age. Hence, the emigration flow can improve the labor market prospects of natives for some specific skills. The rationale behind such finding is due to dynamic complementarity processes between native and emigrant working age population when the shock is originated from emigrant workers. These dynamic processes are consequence of the magnitude of disequilibrium features of the labor market. In contrast, when the shocks are originated from domestic working age population, there is a complex substitutability process between emigrant and native working age population. The net effect from des impulse responses of emigrant and native age dependency shows that the complementarity process between emigrant and native population dominates the substitutability process. Overall, the human capital of emigrants can remove the negative impacts of the increase in Saudi age dependency ratio. These empirical outcomes reflect correctly the stylized facts in Saudi labor market, featured by a complementarity process occurring mostly in the private sector where the emigrants are predominantly employed, but they appear more substitutes than complements in the government sector dominated by Saudi employees.

In addition, in the short-run, the shocks of domestic and emigrant working age populations lead to opposite and unsteady reactions of the economic growth, but in the long-run, both effects turn to be slightly positive. This result means, through the mechanisms of learning by doing and accumulated experiences, that in the long-run the skilled emigrant workers with a fair productivity overcome the unskilled workers with a lower productivity. Our empirical finding corroborates with some previous empirical papers as Dolado et al. (1994) and Ortega and Peri (2009). Furthermore, the emigrants do not disadvantage the native per capita income, and due to their contractual status, the emigrant workers cannot react fully to real wage decreases, compared to the native workforce.

As well, the cumulative impact of demand for emigrant workers on the economic growth is positive; but the cumulative impact of Saudi working age population is associated with a negative economic growth. The net long-run cumulative impact of the two source of population growth on economic growth is positive, due to the predominant contribution of skilled emigrant workers in compared to unskilled. This means that the emigrant workers still occupy many type of jobs either in economic or social activities satisfying the labor demand of families such as private drivers or house-keeping jobs, and labor demand of businesses and companies for skilled and unskilled jobs. These findings make new insights in exploring the relationship between bi-population growth and economic growth inside the neoclassical theory of population.

An important implication of our findings is that the government institutions must improve job opportunities for the native workforce by taking into account the age-structure of the expatriates. Also, such policies should restructure the labor market conditions by matching the skills by age-groups between emigrants and natives. The success of this labor market policy in the private sector depends on smooth and progressive transition from a less productive to a more productive labor force in the entire economy. Furthermore, it depends on the intensity on complementarity and/or substitution between natives and emigrant workers. Another policy implication of our results is that the Saudi economy will gain more by institutionalizing familial grouping for the emigrants. Such a policy would lead to reduce the remittances outflows even if the remittance outflows to GDP ratio is only about 5.8% in 2016, but it requires adjustment and regulation of wages and contractual obligations of the emigrant workers to their Saudi homologues in both private and public sectors.

Overall, the Saudi government, in collaboration with the implied stakeholders, must to find an equilibrium between Saudi workers and emigrant laborers by breaking down the duality of the labor market through the progressive elimination of the sponsorship system, and by increasing the productivity of labor and enhancing the quality of life for all members of the society.

# Appendices: Methodological Notes, Tables and Figures

## Appendix A.

### Appendix A.1. KPSS Test

According to KPSS (1992), their test is a one-sided test because the parameter value of the null hypothesis is for the variance of the random walk, and the stationarity hypothesis is $\sigma_u^2 = 0$. The assumed model is $y_t = \xi t + (r_t + \varepsilon_t)$ with $\varepsilon_t$ a stationary error and where $r_t = r_{t-1} + u_t$ where the $u_t$ are $iid(0, \sigma_u^2)$. The level-stationary and trend-stationarity cases consist of $\xi = 0$ and $\xi \neq 0$, respectively. The null hypothesis is $\sigma_u^2 = 0$ leading to that $y_t$ is $I(0)$, and the alternative hypothesis is $\sigma_u^2 > 0$ leading to that $y_t$ is $I(1)$. As an upper-tail test, the KPSS statistic for testing that $\sigma_u^2 = 0$ is defined by the quantity $\left(T^{-2} \sum_{t=1}^{T} \hat{S}_t^2\right)/s^2(l)$ where $\hat{S}_t = \sum_{j=1}^{t} \hat{\varepsilon}_j$ where $\hat{\varepsilon}_j$ is the residual of a regression of $y_t$ on constant and trend, and $s^2(l)$ is a consistent estimate (i.e. corrected from heteroskedasticity and autocorrelation) of the long-run variance of $\varepsilon_t$ using $\hat{\varepsilon}_j$. This estimate is calculated under the null hypothesis.

KPSS defines $\lambda = \sigma_u^2/\sigma_\varepsilon^2$ measuring the relative importance of the random walk component to the white noise. The series is stationary if $\lambda = 0$, and it has a unit root if $\lambda > 0$. Also, as indicated by KPSS, their assumed model is equivalent to the following ARIMA model $y_t = \xi + y_{t-1} + w_t$ where $w_t = u_t + \Delta \varepsilon_t$. By supposing that $\varepsilon_t$ and $u_t$ are $iid$ and mutually independent and assuming that $w_t$ is autocorrelated of order one, we can write that $w_t = v_t + \theta v_{t-1}$ with $\lambda = -(1+\theta)^2/\theta$, $\lambda \geq 0$ and $|\theta| < 1$. Therefore, we get stationarity when $\theta = -1$, and unit root when $\theta$ is very close to zero (KPSS, 1992, pages 163-164). Under the null hypothesis, the errors $\Delta \varepsilon_t$ are stationary, so the serial correlation of $w_t$ comes from $u_t$. In addition to the usual model of the KPSS test, we use the equivalent model (ARIMA) through the residuals $w_t$ to detect if we get their stationarity or not. This procedure leads to the same results when the $y_t$ have random walk and we retest the stationarity of the first difference of $y_t$.

According to KPSS (page 170, 1992), for a lag truncation parameter $l = 0$ and in absence of autocorrelated errors, the tests have a correct size and power even for small samples, meaning that the asymptotic validity of the tests holds even for small samples when the number of time-observations is between $30 < T < 50$. Using $0 < l \leq 4$, the tests are somewhat less accurate even with more observations. With small samples, the size-distortion becomes more considerable as $l$ is increased, meaning that KPSS test tends to over-reject the true null hypothesis of stationarity mainly for highly autoregressive processes. The choice of $l$ is very important to the test decisions.

### Appendix A.2. Canonical Cointegrating Regression

Generally, we have a scalar $y_t$ and $X_t$ a random vector $(n-1) \times 1$. Let

$$y_t = X_t'\beta + D_{1t}'\gamma_1 + u_{1t} \tag{B1}$$

we consider also that the regressors $X_t$ have the following stochastic processes,

$$\begin{cases} X_t = \Gamma_{21}'D_{1t} + \Gamma_{22}'D_{2t} + \varepsilon_{2t} & (B2) \\ \Delta \varepsilon_{2t} = u_{2t} & (B3) \end{cases}$$

$D_{1t}$ and $D_{2t}$ are deterministic trend regressors and $u_{1t}$ is the unobserved stochastic term. The constant is supposed to be only in $D_{1t}$. It is assumed that $u_t \sim i.i.d.(0, \Sigma)$ with $u_t = (u_{1t}, u'_{2t})$ where $\Sigma$ is a contemporaneous covariance matrix. By defining a right-side long-run covariance matrix $\Lambda$ and full long-run covariance matrix $\Omega$ according to the partition of $u_t$, we have

$$\Sigma = E(u_t u'_t) = \begin{bmatrix} \sigma_{11} & \Sigma'_{12} \\ \Sigma_{21} & \Sigma_{22} \end{bmatrix}, \quad \Lambda = \sum_{\tau=0}^{\infty} E(u_t u'_{t-\tau}) = \begin{bmatrix} \lambda_{11} & \Lambda'_{12} \\ \Lambda_{21} & \Lambda_{22} \end{bmatrix} = [\Lambda_1 \; \Lambda_2],$$

$$\Omega = \sum_{\tau=-\infty}^{\infty} E(u_t u'_{t-\tau}) = \begin{bmatrix} \omega_{11} & \Omega'_{12} \\ \Omega_{21} & \Omega_{22} \end{bmatrix} = \Lambda + \Lambda' - \Sigma \tag{B4}$$

We suppose that the variables $y_t$ and $X_t$ are non-stationary, and that the matrix $\Omega$ has a full rank $n$ and $\Omega_{22}$ is positive definite i.e. non-singular; this implies that the single cointegrating vector is $(1, -\beta')$. Due to the long-run correlation between $u_{1t}$, $\varepsilon_{2t}$ and $\Omega'_{12}$ or $\Lambda'_{12}$, the static OLS estimation of $\beta$ is not valid to make inference on the cointegrating vector (for details, Phillips-Ouliaris 1990, Hansen 1992, Park 1992). If $X_t$ are strictly exogenous regressors i.e. $\Omega'_{12} = 0$ and $\Lambda'_{12} = 0$, then the static OLS estimate has a fully efficient asymptotic normal distribution allowing to use a standard distribution of tests for computing the critical values.

In our empirical work, we use the CCR model (Park 1992) as it employs stationary transformations on the data of $(y_t, X'_t)$. It starts by estimating the unobserved stochastic errors $\hat{u}_t = (\hat{u}_{1t}, \hat{u}'_{2t})$ and constructs corresponding consistent estimates of $\hat{\Omega}$, $\hat{\Lambda} = [\hat{\Lambda}_1 \; \hat{\Lambda}_2]$ and $\hat{\Sigma}$. The data transformation is operated as follows

$$X_t^* = X_t - \left(\hat{\Sigma}^{-1} \hat{\Lambda}_2\right)' \hat{u}_t \tag{B5}$$

$$y_t^* = y_t - \left(\hat{\Sigma}^{-1} \hat{\Lambda}_2 \tilde{\beta} + \left(0, \left(\hat{\Omega}_{22}^{-1}\right)' \hat{\Omega}_{21}\right)\right)' \hat{u}_t \tag{B6}$$

where $\tilde{\beta}$ is an estimate vector $(n-1) \times 1$ of the cointegrating equation parameters, obtained by static OLS that leading to the estimation $\hat{u}_t$. The CCR estimator is asymptotically efficient by applying the OLS method to transformed variables in $y_t^* = X_t^{*'}\beta + D'_{1t}\gamma_1 + u_{1t}^*$, by taking $Z_t^{*'} = (X_t^{*'} \; D'_{1t})$ (Park 1992):

$$\begin{pmatrix} \hat{\beta} \\ \hat{\gamma}_1 \end{pmatrix} = \left(\sum_{t=1}^{T} Z_t^* Z_t^{*'}\right)^{-1} \sum_{t=1}^{T} y_t^* Z_t^* \tag{B7}$$

With this efficient and unbiased estimator, the testing procedures may be operated using standard distributions.

# Appendix B.

**Tables 1.** Correlation, Covariance between CA to GDP and population sharing from 1974 to 2016

| Correlation Covariance | cayr | g_gdp_r | 0-24_ns | 0-24_s | 0-24_to | 25-64_ns | 25-64_s | 25-64_to | 65+_ns | 65+_s | 65+_to |
|---|---|---|---|---|---|---|---|---|---|---|---|
| cayr | 1 0.0252 | 0.0037 | 0.0022 | -0.0036 | -0.0022 | -0.0024 | 0.0035 | 0.0020 | 0.0002 | 0.0001 | 0.0001 |
| g_gdp_r | 0.4848 (0.0011) | 1 0.0023 | 7.7E-06 | -0.0009 | -0.0009 | -1.9E-05 | 0.0009 | 0.0009 | 1.1E-05 | -9.8E-06 | -2.1E-05 |
| 0-24_ns | 0.2392 (0.1271) | 0.0027 (0.9862) | 1 0.0034 | 0.0014 | 0.0025 | -0.0035 | -0.0015 | -0.0026 | 7.8E-05 | 9.1E-05 | 0.0002 |
| 0-24_s | -0.4061 (0.0076) | -0.3299 (0.0328) | 0.4205 (0.0055) | 1 0.0032 | 0.0031 | -0.0013 | -0.0032 | -0.0032 | -6.7E-05 | -3.4E-05 | 6.9E-08 |
| 0-24_to | -0.2273 (0.1476) | -0.3114 (0.0447) | 0.7095 (1.4E-07) | 0.9263 (1.5E-18) | 1 0.0036 | -0.0025 | -0.0032 | -0.0032 | -1.8E-05 | 1.5E-05 | 7.5E-05 |
| 25-64_ns | -0.2519 (0.1075) | -0.0065 (0.9673) | -0.9991 (1.1E-56) | -0.3908 (0.0105) | -0.6879 (4.8E-07) | 1 0.0036 | 0.0014 | 0.0026 | -8.7E-05 | -9.8E-05 | -0.0002 |
| 25-64_s | 0.3987 (0.0089) | 0.3367 (0.0292) | -0.4522 (0.0026) | -0.9988 (5.4E-54) | -0.9397 (3.1E-20) | 0.4236 (0.0052) | 1 0.0032 | 0.0032 | 6.1E-05 | 2.6E-05 | -9.9E-06 |
| 25-64_to | 0.2072 (0.1879) | 0.3117 (0.0444) | -0.7386 (2.3E-08) | -0.9059 (1.6E-16) | -0.9984 (1.8E-51) | 0.7189 (8.2E-08) | 0.9218 (4.6E-18) | 1 0.0037 | 8.9E-06 | -2.5E-05 | -8.9E-05 |
| 65+_ns | 0.3845 (0.0119) | 0.0805 (0.6124) | 0.4717 (0.0016) | -0.4161 (0.0061) | -0.1039 (0.5126) | -0.5084 (0.0006) | 0.3785 (0.0134) | 0.0511 (0.7480) | 1 8.1E-06 | 6.6E-06 | 8.8E-06 |
| 65+_s | 0.2246 (0.1527) | -0.0717 (0.6518) | 0.5508 (0.0002) | -0.2146 (0.1723) | 0.0903 (0.5693) | -0.5773 (6.3E-05) | 0.1664 (0.2922) | -0.1450 (0.3594) | 0.8270 (1.5E-11) | 1 8.0E-06 | 9.8E-06 |
| 65+_to | 0.2502 (0.1100) | -0.1197 (0.4502) | 0.7427 (1.8E-08) | 0.0003 (0.9983) | 0.3385 (0.0283) | -0.7652 (3.6E-09) | -0.0475 (0.7650) | -0.3914 (0.0103) | 0.8373 (4.8E-12) | 0.9391 (3.6E-20) | 1 1.4E-05 |

Notes: The series cayr and g_gdp_r correspond to the current account-to-GDP ratio and real economic growth, respectively. The lower matrix shows the correlation coefficients and their significance level in terms of probability. The upper matrix shows the variance-covariance coefficients. ADR stands for Age Dependency Rate, defined as the ratio between the groups of youth and old (0-24;65+) and the age group 25-64 of working age. The Probability-Values are in parentheses.

| Correlation Covariance | cayr | g_gdp_r | ADR_ns | ADR_s | ADR_to |
|---|---|---|---|---|---|
| cayr | 1 0.0252 | 0.0037 | 0.0073 | -0.0347 | -0.0099 |
| g_gdp_r | 0.4848 (0.0011) | 1 0.0023 | 0.00017 | -0.0076 | -0.0049 |
| ADR_ns | 0.2644 (0.0906) | 0.0202 (0.8990) | 1 0.0303 | 0.0268 | 0.0452 |
| ADR_s | -0.4748 (0.0015) | -0.3433 (0.0260) | 0.3351 (0.0301) | 1 0.2115 | 0.1341 |
| ADR_to | -0.1816 (0.2498) | -0.2947 (0.0581) | 0.7545 (7.9E-09) | 0.8466 (1.6E-12) | 1 0.1186 |

**Table 2.** Descriptive statistics with 42 Observations (1974-2016)

|  | cayr | g_gdp_r | 0-24_ns | 0-24_s | 0-24_to | 25-64_ns | 25-64_s | 25-64_to | 65+_ns | 65+_s | 65+_to |
|---|---|---|---|---|---|---|---|---|---|---|---|
| Mean | 0.06001 | -0.00354 | 0.34443 | 0.63462 | 0.56035 | 0.64585 | 0.33004 | 0.41071 | 0.00972 | 0.03534 | 0.02894 |
| Median | 0.05313 | 0.00271 | 0.32852 | 0.66012 | 0.58223 | 0.66466 | 0.30028 | 0.39002 | 0.00922 | 0.03470 | 0.02790 |
| Std. Dev. | 0.16079 | 0.04881 | 0.05910 | 0.05756 | 0.06068 | 0.06052 | 0.05701 | 0.06205 | 0.00288 | 0.00287 | 0.00374 |
| Skewness | 0.25236 | -1.10866 | 1.44233 | -1.11473 | -0.95745 | -1.57650 | 1.12440 | 0.81459 | 1.04607 | 0.68185 | 1.14416 |
| Kurtosis | 1.77051 | 4.06558 | 4.99893 | 3.44712 | 3.27114 | 5.19635 | 3.31538 | 3.06900 | 3.31071 | 2.86438 | 3.63311 |
| Jarque-Bera (Probability) | 3.09117 (0.21319) | 10.59086 (0.00501) | 21.55476 (0.00002) | 9.04823 (0.01084) | 6.54565 (0.03790) | 25.83946 (0.00000) | 9.02402 (0.01098) | 4.65320 (0.09763) | 7.82884 (0.01995) | 3.28664 (0.19334) | 9.86517 (0.00721) |

**Tables 3.** KPSS stationarity tests

|  | cayr | g_gdp_r | 0-24_ns | 0-24_s | 0-24_to | 25-64_ns | 25-64_s | 25-64_to | 65+_ns | 65+_s | 65+_to |
|---|---|---|---|---|---|---|---|---|---|---|---|
| KPSS LM Stat.[a] | 0.1979 (C, T, 3) | 0.5639 (C, 2) | 0.1707 (C, T, 3) | 0.2101 (C, T, 4.6) | 0.1714 (C, T, 4.5) | 0.1783 (C, T, 3) | 0.2086 (C, T, 4.6) | 0.1734 (C, T, 4) | 0.2196 (C, T, 4.5) | 0.1806 (C, T, 4.3) | 0.1941 (C, T, 4.5) |
| HAC variance[b] | 0.077865 | 0.003791 | 0.006474 | 0.006335 | 0.002599 | 0.007035 | 0.005712 | 0.002139 | 3.64E-05 | 2.52E-05 | 4.05E-05 |

|  | dcayr | dg_gdp_r | d_0-24_ns | d_0-24_s | d_0-24_to | d_25-64_ns | d_25-64_s | d_25-64_to | d_65+_ns | d_65+_s | d_65+_to |
|---|---|---|---|---|---|---|---|---|---|---|---|
| KPSS LM Stat. | 0.2954*** (C, 4.7) | 0.0734*** (C, T, 4) | 0.2154*** (C, 4) | 0.0744*** (C, T, 3.24) | 0.3998** (C, 3.8) | 0.2878*** (C, 3) | 0.0771*** (C, T, 2.9) | 0.1283** (C, T, 4) | 0.0522*** (C, T, 3) | 0.0655*** (C, T, 3.01) | 0.0592*** (C, T, 4.1) |
| HAC variance | 0.007907 | 0.000695 | 0.000602 | 9.51E-05 | 0.000176 | 0.000512 | 8.45E-05 | 0.000125 | 6.76E-07 | 2.08E-06 | 1.42E-06 |

|  | ADR_ns | d(ADR_ns) | ADR_s | d(ADR_s) | ADR_to | d(ADR_to) |
|---|---|---|---|---|---|---|
| KPSS LM Stat. | 0.2059 (C, T, 2) | 0.1817*** (C, 2) | 0.2392 (C, T, 4) | 0.0942*** (C, T, 3) | 0.1868 (C, T, 4.4) | 0.4158*** (C, 5.6) |
| HAC variance | 0.197638 | 0.027490 | 0.030754 | 0.000533 | 0.032179 | 0.003056 |

Notes: [a] Between the parentheses, we have the exogenous variables in the KPSS test equation, Constant C and Linear Trend T. The last number in the parentheses corresponds to Newey-West automatic bandwidth using Quadratic Spectral kernel (Newey and West 1994); when this number is an integer, it means that we use a specified lag truncation parameter. *** and ** stand for 1% and 5% significance levels, respectively. From KPSS (Table 1, 1992): the upper tail percentiles critical values for constant i.e. stationarity around a level $\eta_\mu$ where the residuals are obtained by extracting a mean only from $y$: Critical level: 0.10; 0.05; 0.025; 0.01 Critical value: 0.347; 0.463; 0.574; 0.739. The upper tail critical percentiles values for constant and trend i.e. stationarity around a trend $\eta_\tau$ where the residuals are obtained by extracting a mean and a trend from $y$: Critical level: 0.10; 0.05; 0.025; 0.01. Critical value: 0.119; 0.146; 0.176; 0.216.
[b] HAC variance corresponds to the heteroskedasticity and autocorrelation consistent long-run variance using the quadratic spectral window, which leads to higher rate of consistency (Hobijn, Franses and Ooms 2004).

**Table 4.1.** Residual correlation matrix

|  | $res\_dcay_t$ | $res\_ds\_adr_t$ | $res\_dns\_adr_t$ | $res\_dg\_gdp_t$ |
|---|---|---|---|---|
| $res\_dcay_t$ | 1.00000 | 0.00556 | 0.39096 | 0.33449 |
| $res\_ds\_adr_t$ |  | 1.00000 | $-0.31024$ | $-0.02763$ |
| $res\_dns\_adr_t$ |  |  | 1.00000 | 0.38270 |
| $res\_dg\_gdp_t$ |  |  |  | 1.00000 |

**Table 4.2.** Residual variance-covariance matrix

|  | $res\_dcay_t$ | $res\_ds\_adr_t$ | $res\_dns\_adr_t$ | $res\_dg\_gdp_t$ |
|---|---|---|---|---|
| $res\_dcay_t$ | 0.00919 | $8.11E-06$ | 0.00033 | 0.00131 |
| $res\_ds\_adr_t$ |  | 0.00023 | $-4.21E-05$ | $-1.72E-05$ |
| $res\_dns\_adr_t$ |  |  | $7.96E-05$ | 0.00014 |
| $res\_dg\_gdp_t$ |  |  |  | 0.00167 |

**Table 5.** Structural variance decomposition

|  | TIT | | SWP | | DEL | | DOS | |
|---|---|---|---|---|---|---|---|---|
|  | h=10 | h=25 | h=10 | h=25 | h=10 | h=25 | h=10 | h=25 |
| DCAY | 61.40 | 49.40 | 3.40 | 7.62 | 32.45 | 38.01 | 2.75 | 4.97 |
| DS_ADR | 0.03 | 0.04 | 23.05 | 22.67 | 72.65 | 69.41 | 4.27 | 7.88 |
| DNS_ADR | 2.69 | 2.37 | 15.05 | 21.22 | 69.15 | 63.73 | 13.11 | 12.68 |
| DG_GDP | 13.43 | 13.32 | 2.35 | 2.49 | 5.24 | 5.73 | 78.98 | 78.46 |

Note: Variation in the row variable explained by shocks in column variable.
The number are in percent for 10 periods 25 periods (generation) ahead.

# Appendix C.

**Figures C.0.1.** Bi-Population Pyramid by Census in Saudi Arabian Economy from 1974 to 2016

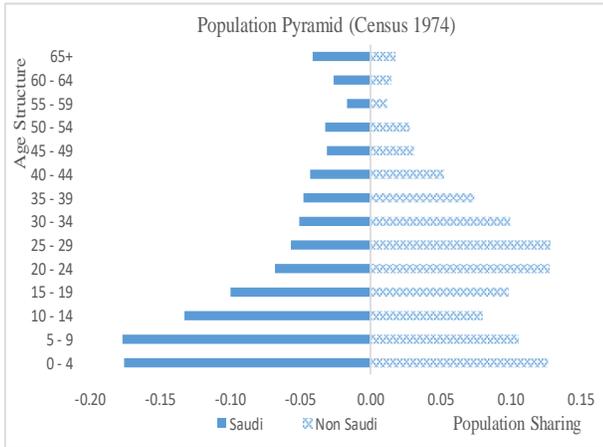
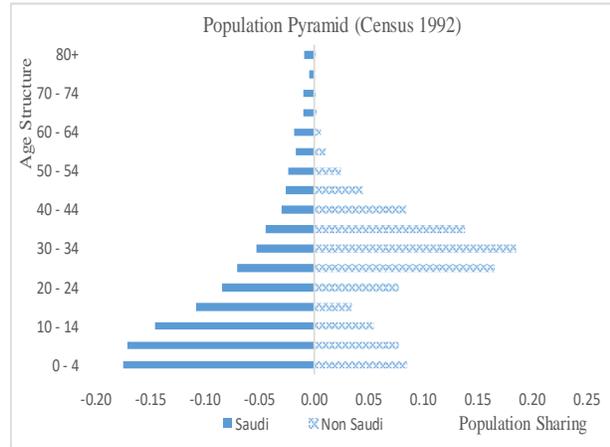
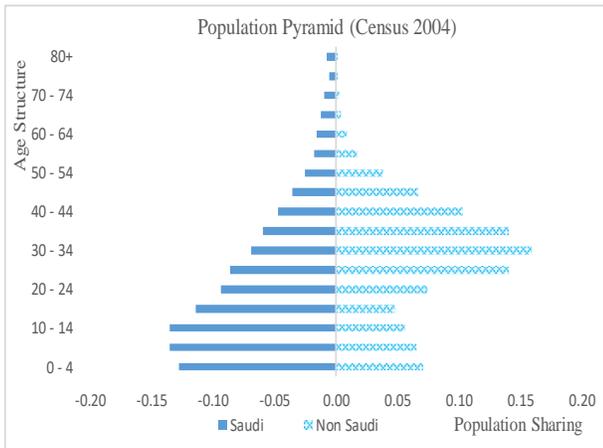
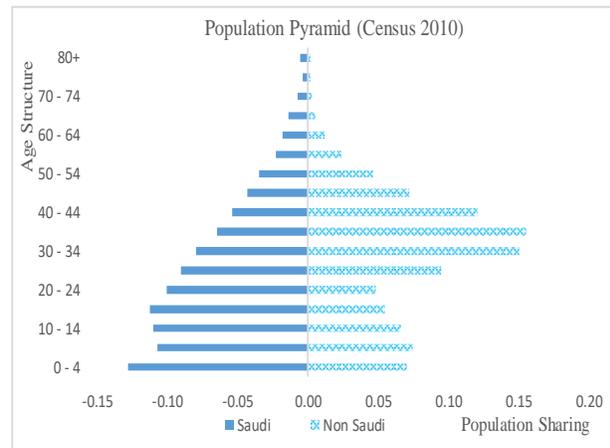
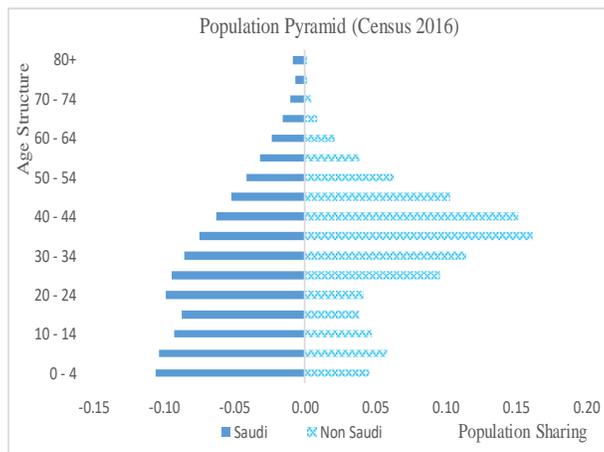

**Figures C.0.2.** Bi-Age Structure in Saudi Arabian Economy from 1974 to 2016

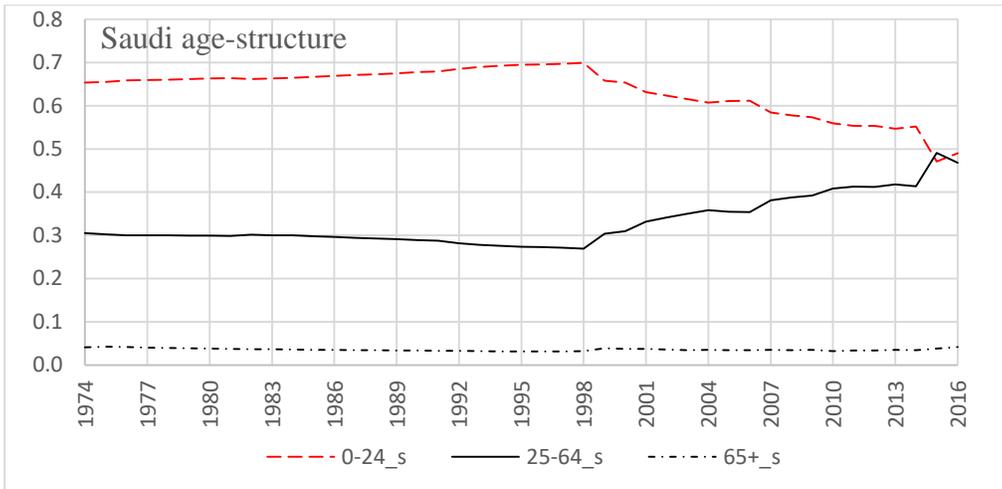

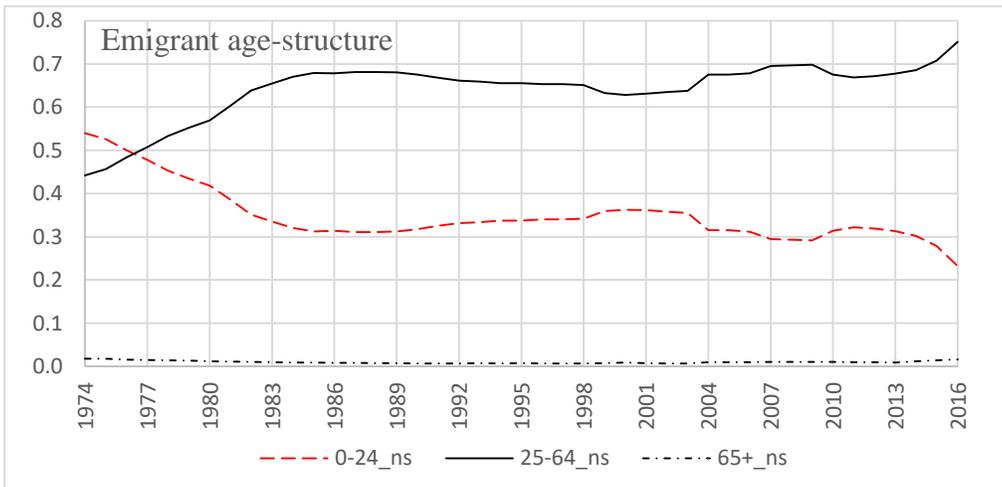

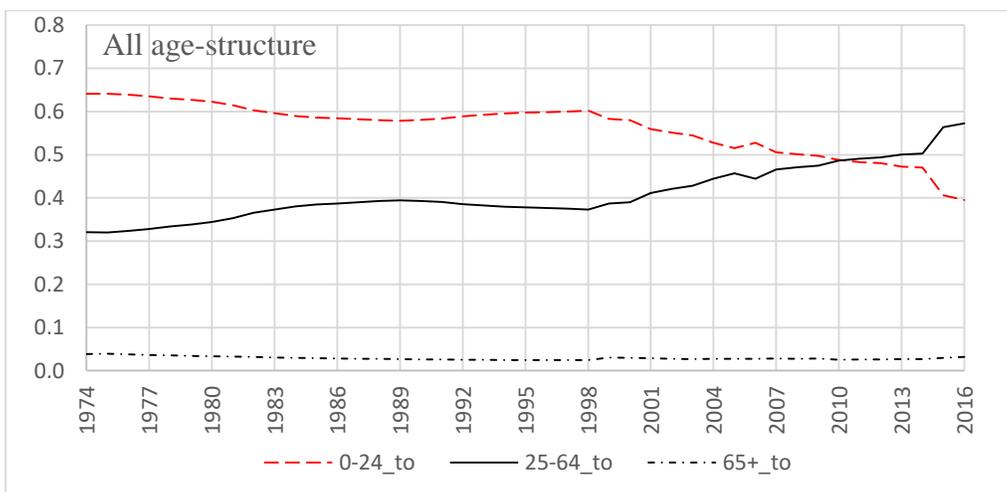

**Figures C.1.** Impulse responses functions of Current account-to-GDP ratio [26]

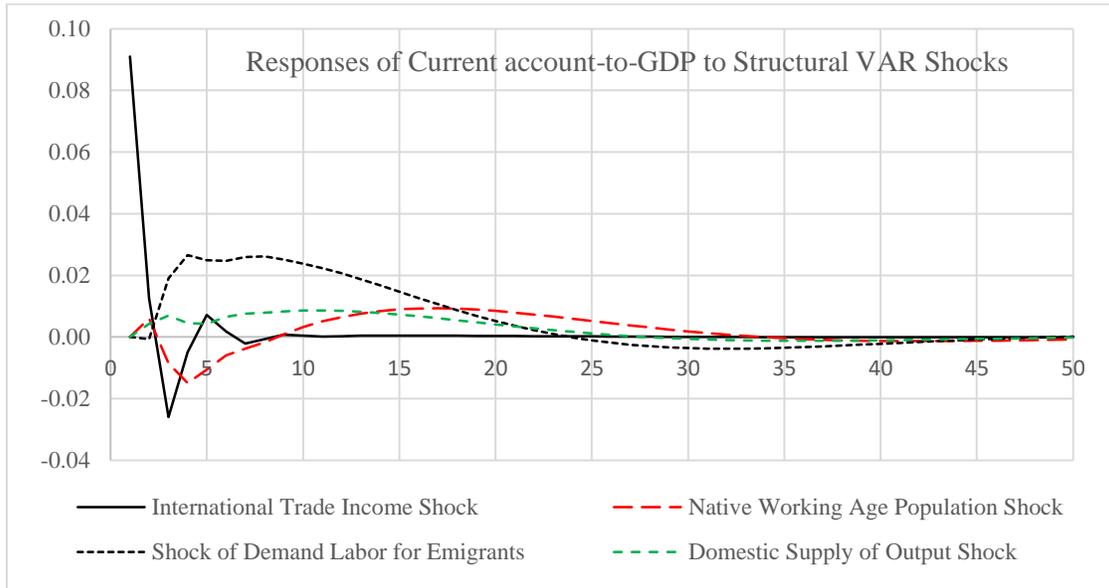

**Figures C.2.** Impulse responses functions of Saudi Age dependency rate

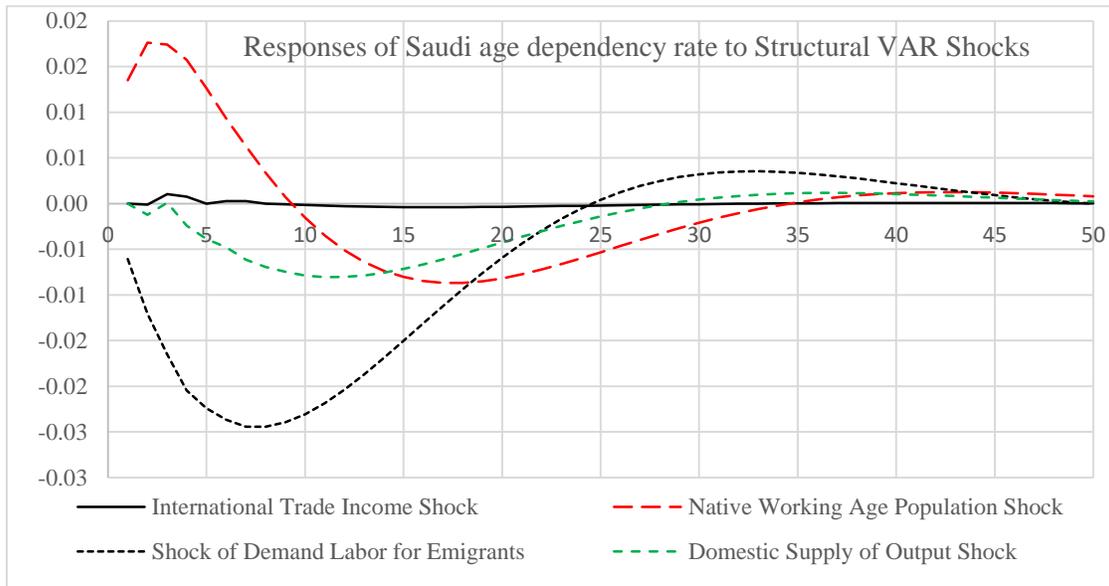

---

[26] The confidence intervals of the impulse response functions, based on structural decomposition, are computed using Monte-Carlo simulations with 5000 times replications. The solid line indicates the impulse responses, the dashed lines display five standard error confidence bands. As we estimate a stationary VAR model in first differences, the impulse responses are consistent at long horizons; thus, the stationarity means that the impulse responses do tend to zero as the horizon increases.

**Figures C.3.** Impulse responses functions of Emigrant Age dependency rate

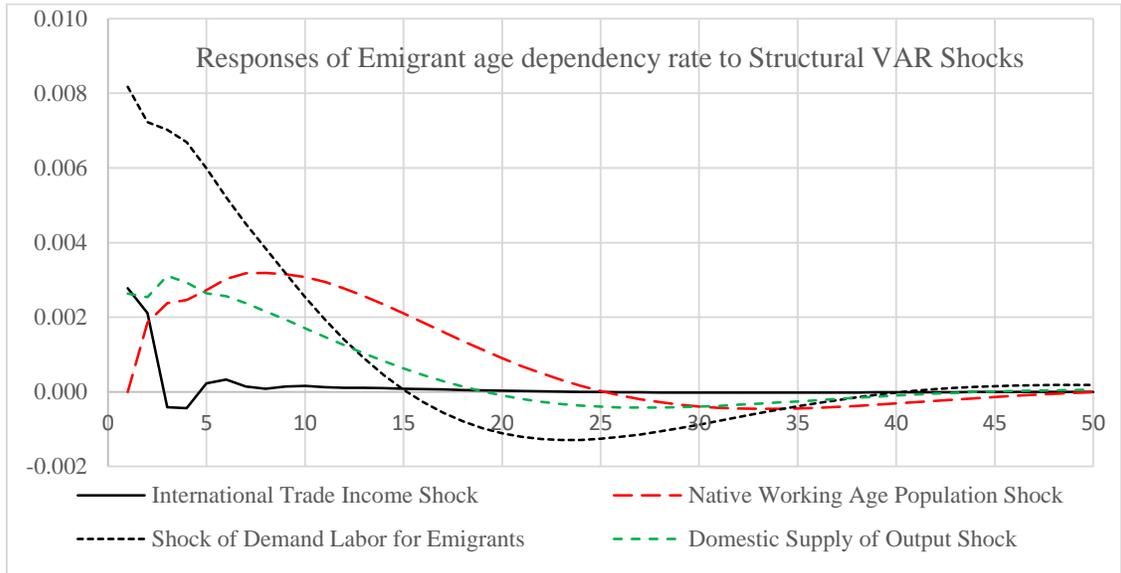

**Figures C.4.** Impulse responses functions of Economic growth

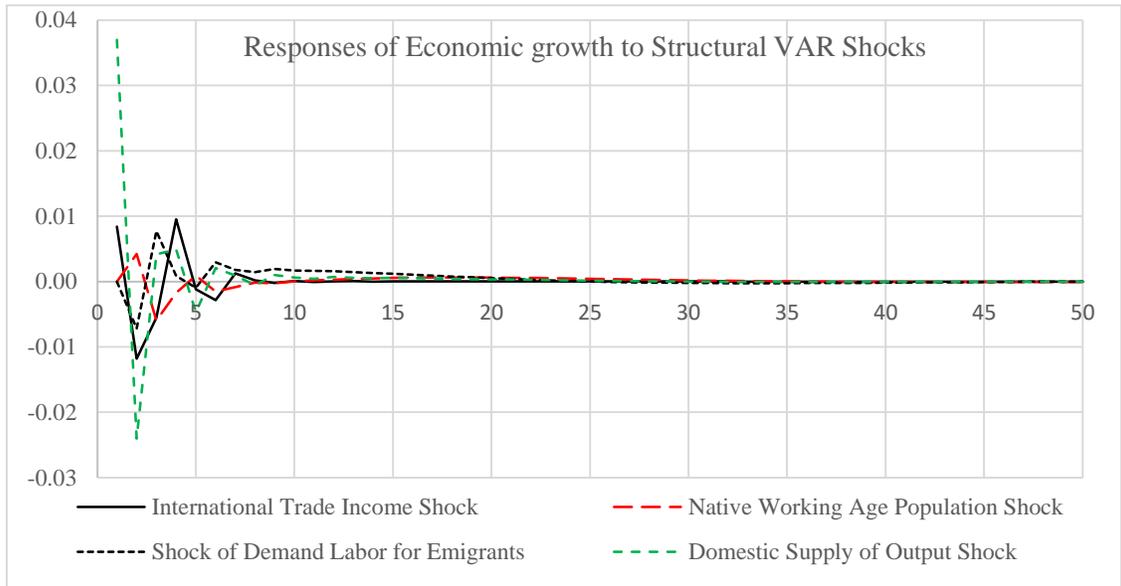